\documentclass[11pt]{amsart}
\usepackage[a4paper, total={6in, 8in}]{geometry}
\usepackage{epsfig}
\usepackage{amsmath}
\usepackage{amssymb}
\usepackage{bbold}
\PassOptionsToPackage{hyphens}{url}\usepackage{hyperref}

\usepackage[natbibapa,nodoi]{apacite}
\usepackage[labelfont=bf]{caption}
\usepackage{setspace}
\usepackage{subcaption}
\usepackage{cleveref}
\usepackage{lscape}
\usepackage{verbatim}
\usepackage{multirow}
\usepackage{bm}
\usepackage{array} 
\usepackage{color}
\newtheorem{theorem}{Theorem}[section]
\newtheorem{lemma}[theorem]{Lemma}
\newtheorem{proposition}[theorem]{Proposition}

\usepackage{graphicx}
\usepackage{wrapfig}
\DeclareMathOperator{\Tr}{Tr}
\newcommand{\vb}{\vspace{3.2mm}}

\allowdisplaybreaks
\usepackage[normalem]{ulem}

\begin{document}

\title[A relative approach to opinion formation]{A relative approach to opinion formation}

\author[Chan, Duivenvoorden, Flache, and Mandjes]{ Kit Ming Danny Chan$^{1,2}$, Robert Duivenvoorden$^{2}$,\\ Andreas Flache$^{3}$, and Michel Mandjes$^{1}$}

\begin{abstract} 
Formal models of opinion formation commonly represent an individual's opinion by a value on a fixed opinion interval. We propose an alternative modeling method wherein interpretation is \textit{only} provided to the relative positions of opinions vis-à-vis each other. This method is then considered in a similar setting as the discrete-time Altafini model (an extension of the well-known DeGroot model), but with more general influence weights. Even in a linear framework, the model can describe, in the long run, polarization, dynamics with a periodic pattern, and (modulus) consensus formation. In addition, in our alternative approach  
key characteristics of the opinion dynamic can be derived from real-valued square matrices of influence weights, which immediately allows one to transfer matrix theory insights to the field of opinion formation dynamics under more relaxed conditions than in the DeGroot or discrete-time Altafini models. A few specific themes are covered: (i)~We demonstrate how stable patterns in relative opinion dynamics are identified which are hidden when opinions are considered in an absolute opinion framework. (ii)~For the two-agent case, we provide an exhaustive closed-form description of the relative opinion model's dynamic in the long run. (iii)~We explore group dynamics analytically, in particular providing a non-trivial condition under which a subgroup's asymptotic behavior carries over to the entire population.

\vb


\vb

\noindent
$^{1}$Korteweg-de Vries Institute, University of Amsterdam, the Netherlands
    
\noindent  $^{2}$Transtrend {\sc bv}, Rotterdam, the Netherlands
    
\noindent $^{3}$Department of Sociology, University of Groningen, the Netherlands
    
\noindent    \vspace{0.1cm}
    
\noindent Corresponding author: Kit Ming Danny Chan ({\tt k.m.d.chan@uva.nl}).

\vspace{0.5cm}

\noindent \textit{This article has been accepted for publication in The Journal of Mathematical Sociology, published by Taylor \& Francis.}

\end{abstract}

\maketitle

\newpage 

\section{Introduction}\label{section_intro}
Opinion formation processes play a prominent role in societies. Recent examples are the polarization of opinions regarding {\sc covid}-19 measures or the demonstrations and political debates on racial discrimination. Some of these processes correspond to short-term dynamics, such as the fluctuating views on the use of wearing face masks to combat {\sc covid}-19. Other examples reflect more long-term opinion formation processes like the cycles in fashion industry \citep{Aspers2013}, where the reintroduction of colors seems to follow a periodic pattern extending over many years. 

The underlying dynamics behind opinion formation are highly complex due to the interdependence of interconnected individuals influencing each other, directly or indirectly, in various ways \citep{Flache2017, Friedkin1990, Kozitsin2020, Lee2020}. There is a large body of empirical studies \citep{Bramson2016} of opinion dynamics in society, among which many are performed through surveys. However, despite important theoretical advances, thus far the understanding of how individual-level opinion changes translate into large-scale phenomena, such as polarization or consensus formation, has remained limited. For empirical studies it is a complicating factor that the number of agents involved can be large, making data collection challenging. But even when large amounts of data are available, a theoretical modeling framework of opinion formation is desirable, so as to facilitate a sound interpretation of the data. A long standing challenge has been to explain phenomena such as  polarization, consensus formation, and periodicity relying on relatively compact and transparent models.

Over the past decades great advances have been made and a large variety of opinion formation models have been proposed. Linear models were studied assuming that agents are willing to move their opinions closer to those of sources of influence in a network. A seminal model of this type is the DeGroot model \citep{DeGroot} where opinions develop according to a weighted averaging principle. Here, agents typically reach a consensus when the population is ``well-connected'' and ``aperiodic''\footnote{``Well-connected" and ``aperiodic'' in the DeGroot model and the discrete-time Altafini model \citep{Altafini2012,Altafini2013,Liu2017} means that the driving matrix is irreducible and not periodic respectively.}. In further exploring richer dynamics, researchers introduced more complex, sometimes nonlinear opinion formation models, such as \citet{Hegselmann2002, Friedkin1990, Flache2011b}. Some of these nonlinear models can be investigated analytically, such as the influential Friedkin-Johnsen model \citep{Friedkin1990}. Other models are mainly analyzed through simulations, such as the Bounded Confidence model \citep{Krause1997, Deffuant2000} and other agent-based models \citep{Flache2017}. While most of these models generate opinion dynamics which lead to convergence of the set of opinions in a population to a bounded finite range, some models added assumptions that also induce persistent divergence, periodicity or shift of opinions over time \citep{Friedkin1990,Flache2017}. This applies in particular to models which added the possibility that agents can be ``negatively'' influenced, seeking to distance their opinions from those of negatively evaluated sources -- the so-called boomerang effect \citep{Altafini2013,Flache2011b}. 
\vb

One outstanding research problem in opinion formation models concerns finding a simple framework in which the main large-scale phenomena (including polarization, consensus formation, and periodicity) can be explained in the long run, explicitly covering situations in which the population is ``well-connected" and ``aperiodic''. The underlying model dynamics would preferably be linear, thus allowing an explicit analysis relying on matrix-theoretic tools. Having such a relatively elementary framework would clearly be highly appealing, as  it would allow one to reason about the major opinion formation phenomena through a single transparent, but highly general, model. 

A second open problem relates to the interpretation of model outcomes for opinion formation processes which play out simultaneously in long-term and short-term time scales. Especially linear models based on the DeGroot framework either generate consensus or persistent divergence of opinions from each other in the long run. This can obscure that there may be robust patterns of fragmentation in a population that persist over long time periods even when the overall distribution\footnote{In this paper the {\it distribution} of opinions refers to the spread of opinions of the agents considered, i.e., it is {\it not} meant as a probability distribution.} of opinions moves towards consensus, or shifts to other intervals in an opinion spectrum, or keeps diverging. For example, an opinion distribution exhibiting consensus when observed from one historical frame of reference, may seem to reveal strong polarization from another frame of reference---albeit potentially within a shifted or much smaller range of socially acceptable views. 
\vb

To illustrate the possibility of shifting opinion ranges highlighted in the previous paragraph, consider changes in opinions in society regarding racial discrimination between the 18th century and now. Where at one point in history a large portion of the non-enslaved population held the opinion that enslaving members of racial minorities was acceptable behavior while a minority fiercely disagreed, about two centuries later outright rejection of this view thankfully has become the clear consensus. However, to the extent that differences in opinions about existence and justification of social, economic or judicial inequalities between racial groups can be seen at the same opinion scale as acceptance or rejection of slavery, events like the recent rise of the {\it Black Lives Matter} movement illustrate how strong polarization can occur within a shifted range of existing views that may appear as reflecting near-consensus (about rejecting slavery) when compared to an opinion distribution from a different historical period.

The shift of opinion ranges over time or persistent divergence of opinions can occur particularly when processes of distancing from negatively evaluated opinion sources can generate boomerang effects. There is an ongoing debate in the literature about the extent to which such boomerang effects actually occur in empirical social influence. For example, in a controlled laboratory experiment  \citet{Takacs2016} found no convincing empirical support for a boomerang effect. However, as mentioned in \citet{Flache2017}, this is insufficient evidence to exclude boomerang effects in situations where they would be expected on theoretical grounds, such as for strong emotional content \citep{Sobkowicz2012, Sobkowicz2015}, high ego-involvement, or strong antagonistic group identities \citep{Huet2010}. Accordingly, empirical field studies have aimed to test the possibility of boomerang effects and find at least some support for opinion shifts away from the position of a source of influence, possibly caused by a boomerang effect \citep{Liu2015, Bail2018, Levendusky2013, Kozitsin2021}. 

The theoretical significance of opinion dynamics involving boomerang effects led researchers \citep{Altafini2013, Proskurnikov2016, Liu2017, Shi2019, Proskurnikov2020} to explore their implications for opinion dynamics. An example of such a model with linear influence dynamics is the discrete-time Altafini model (an extension of the DeGroot model) which captures in the long run a specific type of polarization: modulus consensus formation\footnote{In the long run, i.e. as time proceeds indefinitely, an agent's opinion evolves to either $x$ or $-x$ with $x \in \mathbb{R}_+$.}. Still, in a ``well-connected'' population the DeGroot and discrete-time Altafini model are limited to describing (modulus) consensus formation in the long run; general polarization, persistently shifting opinion ranges or periodic evolution of opinions remain unexplained\footnote{Note that the DeGroot model and the discrete-time Altafini model can asymptotically describe general polarization and periodic evolution of opinions when the driving matrix is reducible or periodic.}. 
\vb

In this paper we contribute to addressing both research problems highlighted above. We propose a simple and transparent mathematical framework that can help explain polarization, consensus formation, and periodicity in the long run in a ``well-connected'' and ``aperiodic'' population, even when such patterns cannot be observed within limited sections of an opinion spectrum. The key innovation of the framework is the idea of \textit{only} modeling the relative positions of opinions vis-à-vis each other, which we will call \textit{relative opinions}. This approach provides an alternative viewpoint on opinion modeling and allows one to contemplate in a general but also compact framework about opinion dynamics. We will explain and motivate relative opinions in greater detail in Section \ref{section_motivation}. From a technical point of view, we will model relative opinions in a setting similar to the discrete-time Altafini model, but with more general influence weights\footnote{We consider more general influence weights in the sense that any real-valued square matrix can be considered as driving matrix of the system. The discrete-time Altafini model assumes that the driving matrix $A$ can be time-dependent, in this paper we assume that $A$ does not change over time. We leave the time-dependent variant for further research.}; the \textit{relative opinion model}. We will show that it is especially suitable for modeling agents who can adjust their opinion both positively and negatively after interaction with agents, while still remaining in a linear framework. As a result, the model succeeds in describing rather complex dynamics while working with a relatively simple underlying mechanism. 

We believe that the consideration of relative opinions is also of significance in aiding the interpretation of results found in empirical research. Longer term patterns may be challenging to identify for studies focusing on changes observed in distributions of opinions measured with fixed attitude scales, which are typically used by survey instruments \citep{Bramson2016}. 

\vb

In Section \ref{section_motivation} we explain and motivate the idea of modeling relative opinions in greater detail. Then, in Section \ref{section_model} the relative opinion model is introduced. The dynamics of the model are of the same complexity as in the DeGroot model, i.e., in essence described by a linear map. Section \ref{section_analysis} discusses structural properties that are used in the analysis of later sections. In addition, the model's asymptotic (long term, that is) behavior is focused on.
For the specific case of two agents, we present an exhaustive closed-form description of the relative opinions' asymptotic behavior, in particular showing how polarization (in the sense of persistent disagreement), consensus formation, and periodicity indeed arise. At the methodological level, no heavy machinery is used: elementary techniques from matrix algebra and Markov chain theory suffice.

Section \ref{sect_groups} presents an analysis of two clusters of agents or ``groups" \citep{Altafini2012,EGER2016,Proskurnikov2016}  and the corresponding asymptotic opinion formation dynamics. We conclude with a discussion which includes suggestions for further work in this area.

The complete Matlab code and datasets used in generating the figures in this article are publicly available through 
\url{https://www.comses.net/codebase-release/14f7267a-e6e5-493c-9fc5-954a1d37f928/}.

\section{Explanation and motivation of modeling relative opinions}\label{section_motivation}

In order to explain \textit{relative opinions} in the context of the current literature on opinion formation models we first discuss the custom of modeling opinions on a bounded range $[a,b]$ with $a,b \in \mathbb{R}$ and $a < b$, which we call the \textit{absolute opinion} framework. The assumption of a bounded range is in various models not specified as an explicit component of the model's definition, instead it follows indirectly from the considered influence parameters. As an illustration, in the DeGroot model the sum of the influence weights exhibited on any agent must equal $1$, while a similar condition holds for the discrete-time Altafini model\footnote{Here, the sum of \textit{absolute values} of the possibly negative influence weights exhibited on any agent must equal $1$.} \citep{Liu2017}. In these examples, the considered influence parameters in fact restrict opinions to only evolve (modulus) towards each other. When more general negative influences between agents\footnote{More general with respect to the influences considered in the discrete-time Altafini model.} are considered as well, opinions may evolve \textit{unboundedly away} from each other. In order to ensure that opinions stay within bounds, various models have been studied with modifications of the influence parameters. Examples of modifications are smoothing \citep{Flache2011b} and truncating functions \citep{feliciani2017}. An important drawback of models with such restrictions or modifications of influence parameters is that they generally lack convincing empirical backing. For example, to our best understanding no convincing empirical support has been provided for the claim that the sum of absolute influence weights exhibited on any agent equals 1, as imposed in the discrete-time Altafini model. 

An alternative remedy to the problem that ``opinions may evolve unboundedly'' is the modeling of relative opinions, which we will explain here. It should be noted that the remedy arises as a by-product of modeling relative opinions; other empirically grounded motivations for relative opinions are discussed further in this section. In the tradition of the bounded range $[a,b]$, the boundaries $a$ and $b$ act in fact as two reference points. Indeed, an opinion $y \in \mathbb{R}$ with $a \leq y \leq b$ is interpreted as an opinion that is $y-a$ and $b-y$ ``away'' from the boundaries $a$ and $b$ respectively. The bounded range is a modeling assumption, other choices may provide different insights into opinions dynamics. Hence, alternatively we can choose the value $0$ as our \textit{only} reference point, which also provides a convenient interpretation of ``positive'' and ``negative'' opinions: positive values are interpreted as ``positive'' opinions and vice versa. The essential difference with a bounded range is the actual content or interpretation we seek to model. The distance of an opinion $y_1 \in \mathbb{R}$ to the reference point is now $y_1-0$, stand-alone we do not grant interpretation to this distance except for its sign. Instead, we only grant interpretation to the \textit{relative} distance of $y_1$ with respect to another agent, say with opinion $y_2 \in \mathbb{R}$, namely the fraction $\frac{y_1}{y_2}$. Summarizing, an agent 1's opinion $y_1$ only reveals the side of the opinion spectrum it is on, ``positive'' or ``negative'', and nothing about the magnitude of its opinion. Only when agent 1's opinion is viewed with respect to an agent 2's opinion, one gains a sense of the magnitude of agent 1's opinion, namely agent 1's opinion is a $\frac{y_1}{y_2}$ fraction of agent 2's opinion. The complete interpretation of $y_1$ follows from comparing it with all $N>1$ agents in the population: the fractions $\frac{y_1}{y_2}, \ldots, \frac{y_1}{y_N}$, note that these are $N-1$ terms instead of $N$. We thus only model the relative position of agent's opinion vis-à-vis each other and (indirectly) the reference point $0$ -- \textit{relative opinions} -- which clearly lacks the notion of boundaries.

The idea of relative opinions can also be viewed from a slightly different angle. An interpretation is that we are following and perceiving the population's opinion through the eyes of a single agent, the Observer. The Observer perceives the distance between its own opinion and zero as the ``unit distance'' and only observes other agents' opinion with respect to it. Interesting is that every Observer perceives relative distances between all agents identically (up to a sign)\footnote{An Observer 1 perceives the opinion of agents 2 and 3 as $\frac{y_2}{y_1}$ and $\frac{y_3}{y_1}$, their relative distance is $\frac{y_2}{y_1} / \frac{y_3}{y_1} = \frac{y_2}{y_3}$, the latter term is not dependent on the Observer's opinion $y_1$.}. In the field of economics/finance, changing Observer's view is similar to the concept of changing the \textit{numéraire}.

Considering relative opinions has advantages in various settings. The relative opinion framework is suitable for capturing relative position shifts in agents' opinion, including consensus formation and polarization. In particular it can describe opinions that shift from the positive side of the opinion spectrum to the negative and vice versa. An empirical example of such a shift is the change of British attitudes about homosexual relations \citep{Chattoe2014}: in 1987 around 75\% of the sample of the British population studied in the British Social Attitude survey were negative on homosexual relations. This decreased to 30\% by 2010. Another example is the decrease of whites’ in the U.S. agreeing to home sellers’ discrimination from around 65\% in 1972 to below 30\% in 2008 \citep{Bobo2012}.

An especially advantageous setting for relative opinions is where opinion boundaries are \textit{unnecessary} to define beforehand. When a bounded opinion range is defined too narrowly, the model may already exclude (unexpected) phenomena by the choice of opinion range. As an illustration, in the questionnaire of the 1972 U.S. General Social Survey \citep{GSS_Questions_1972} participants were asked ``Do you think Negroes should have as good a chance as white people to get any kind of job, or do you think white people should have the first chance at any kind of job?’’. Not only the wording of the question but also the range of answering options used in that survey reveals how opinions in U.S. society have dramatically shifted since then. In 1972, participants could choose from the answers \textit{1. As good a chance}, \textit{2. White people first}, \textit{8. Don’t know}. Hypothetically, a modeler in 1972 that applied this opinion scale would not be able to describe opinions evolving to the opinion \textit{Black people first}. In general, it is arguable that the range of offered survey answers to participants at any period in time is defined by the range of acceptable opinions at that moment. So, when one wants to model longer term opinion dynamics one cannot restrict opinion ranges to the values used in a survey at a given point in time. In the longer time-scale it can be challenging to define the outermost boundaries without excluding phenomena since it requires a certain vision of the future. An example in political sciences where the relative approach seamlessly fits in is the study of the range of acceptable political thoughts, the so-called Overton Window \citep{Dustin2019}. In the perspective, what is ``acceptable'' at a given moment in history is mostly dependent on the position of opinions vis-à-vis each other, which is precisely what the relative opinion framework focuses on. 

A drawback of a bounded opinion range is that it only allows modelers to study \textit{subsets of influence parameters} which ensure that opinions remain within the defined boundaries. To our best understanding these restrictions on influences between agents are generally not supported by empirical evidence. In contrast to the bounded opinion range, the relative opinion framework is not restricted to any set of influence parameters. Where in the bounded opinion range agents influence each other less and less as opinions move closer to a boundary, influences between agents in the relative opinion framework can remain unchanged. From an Observer's point of view, influences  can remain intact regardless of the position of its own opinion, since its opinion has, by definition, no meaning in isolation (except for its sign).

\section{Model}\label{section_model}
\subsection{Model description}\label{subsection_model_description}
We propose a model where the relative opinion framework is considered in a similar setting as in the discrete-time Altafini model \citep{Liu2017}, but with more general influence weights. The object of study is a vector $\boldsymbol{y} \in \mathbb{R}^N$ with entries $y_1, \ldots, y_N$, with $N > 1$ being the number of agents in the population. In models such as the DeGroot model or the discrete-time Altafini model the stand-alone entries of $\boldsymbol{y}$ typically describe the agents' opinion. We choose an \textit{alternative} interpretation of $\boldsymbol{y}$ where the actual content of $\boldsymbol{y}$ lies in the sign and in the relative magnitude of the entries as described by\footnote{We consider a zero entry of $\boldsymbol{y}$ as a singular situation where the agent with zero opinion is excluded from the definition of relative opinions (until this opinion becomes non-zero due to influences by other agents).\label{foot_singular}} \begin{equation}\label{eq_relative}
    \frac{y_i}{y_j}, \,\,\, i,j = 1, \ldots, N,
\end{equation} which we call \textit{relative opinions}. Interestingly, relative opinions can be sufficiently represented by $N-1$ terms, instead of the $N$-terms present in $\boldsymbol{y}$. For example consider the $N-1$ terms where entries of $\boldsymbol{y}$ are denoted in units of agent $1$'s opinion (the Observer see Section \ref{section_motivation}): $\frac{y_2}{y_1}, \ldots, \frac{y_N}{y_1}$. Clearly, the relative opinions vis-à-vis agent $1$ follows immediately. Also, the fraction $\frac{y_i}{y_j}$ which is the relative opinion between agents $i$ and $j$ ($i,j = 2, \ldots, N$) can be obtained by dividing the terms $\frac{y_i}{y_1}$ with $\frac{y_j}{y_1}$.

Thus, we use an $N$-dimensional vector, while really \textit{only} being interested in relative opinions that can be described by $N-1$ terms. This difference in dimension hints at the absence of an 1-to-1 mapping between the set of all $\boldsymbol{y} \in \mathbb{R}^N$ and the set of all relative opinions that can be constructed from $N$ agents. Indeed, any vector $\gamma \, \boldsymbol{y} \in \mathbb{R}^N$ with scalar $\gamma > 0$, represents the identical relative opinions as $\boldsymbol{y} \in \mathbb{R}^N$ . This can be easily seen by noting that $\gamma$ appears both in the numerator and denominator of expression (\ref{eq_relative}). The multiplicative scalar $\gamma$ is restricted to be positive in order to preserve the side of an agent's opinion on the opinion spectrum. A vector $\boldsymbol{y} \in \mathbb{R}^N$ is thus always one of many representations of the underlying relative opinion; we therefore call these vectors \textit{representative} opinion vectors.

We further elucidate representative opinion vectors with a numerical description of a three agent society with a \textit{mildly negative}, a \textit{mildly positive} opinion and a \textit{strongly positive} opinion, which might for example be described by the column opinion vector $(-1, 1, 2)^\top$ (with ${\boldsymbol x}^\top$ the transpose of the vector ${\boldsymbol x}$). From a relative point of view, the third agent's positive opinion is twice as large in magnitude as the first agent's negative opinion and of the opposite sign, while the second agent's opinion is half the magnitude of the third agent's opinion and of the same positive sign. Note now that the opinion vector in which all opinions are scaled by a factor $2$, that is the vector $(-2, 2, 4)^\top$, still yields the same relative description.

We are ready to define the equivalence relation that relates identical representative opinion vectors. With ${\boldsymbol a}, {\boldsymbol b} \in {\mathbb R}^N$,
\[\mbox{${\boldsymbol a}$ and ${\boldsymbol b}$ are equivalent if a scalar $\gamma > 0$ exists such that ${\boldsymbol a} = \gamma \,{\boldsymbol b}$}.\]
We write: ${\boldsymbol a} \equiv {\boldsymbol b}$.

Under the equivalence relation, representative opinion vectors develop according to an iterated weighted averaging principle that builds on, but also differs from, the discrete-time Altafini model. In contrast to the discrete-time Altafini model we do not restrict the sum of absolute values of the possibly negative influence weights exhibited on any agent to $1$. Instead, we consider \textit{general} influence weights between agents, hence, to a certain extent generalizing the discrete-time Altafini model further. In the relative opinion model each agent $i$ is tied to another agent $j$ by a susceptibility weight $a_{ij} \in {\mathbb R}$. The opinion of agent $i$ is repelled by the opinion of agent $j$ when the weight $a_{ij}$ is negative\footnote{We build further on the literature that explores the implications in those situations where the boomerang effect does apply, see Section \ref{section_intro}.}, attracted to the opinion of $j$ when the weight is positive, and is completely unaffected by the opinion of $j$ when the weight equals $0$. Thus, each agent's way of processing the opinions of all agents in the population can be described by a (row) vector of susceptibility weights. The susceptibility weights of the entire population can therefore be summarized in a square updating matrix $A = \{a_{ij}\}_{1 \leq i,j \leq N}$, with $a_{ij} \in {\mathbb R}$. Given such an updating matrix $A$ and an initial representative opinion vector ${\boldsymbol y}^{(0)} \in \mathbb{R}^N$, the development of the representative opinion vector ${\boldsymbol y}^{(t)}$ at time $t = 1, 2, 3, \ldots$ is described, in the context of \textit{relative opinions}, by the difference equation
\begin{equation} \label{eq_ROM_description}
{\boldsymbol y}^{(t+1)} = A \,{\boldsymbol y}^{(t)},
\end{equation}
with $A$ a real valued square $N \times N$ matrix and ${\boldsymbol y}^{(t)}$ a real valued $N$-dimensional representative opinion vector. Although the model is defined for general $A$ we will focus in this paper on $A$ with $\sum_{j} a_{ij} > 0$, $i = 1, 
\ldots, N$, which we will explain at the end of Subsection \ref{subsection_attentiveness_persistence}. It is worth to emphasize that Equation (\ref{eq_ROM_description}) describes \textit{only} the development of the relative positions of agents' opinion, as in expression (\ref{eq_relative}). Thus, although the formulation of the difference equation is highly similar to, for example, the DeGroot or the discrete-time Altafini model, the interpretation is completely different. 

\subsection{Fixing a representation}\label{subsection_fix_representation}
Some representations of relative opinion dynamics would help the reader better in interpreting the patterns than others. For example, representing the development of $\boldsymbol{y}^{(t)}$ ``as is'' -- as described by Equation (\ref{eq_ROM_description}) -- certainly contains all the relevant information required to obtain the fractions in expression (\ref{eq_relative}), however, not necessarily in a compact or insightful manner. Since ${\boldsymbol y}^{(t)}$ in Equation (\ref{eq_ROM_description}) are representative opinion vectors, and thus considered under the equivalence relation, it represents the same state of opinions as $\gamma \,{\boldsymbol y}^{(t)}$ for any $\gamma > 0$. Hence, there is the freedom to transform a representative opinion vector to an equivalent one. These transformations can be seen as choosing different representations for the same state of relative opinions, with the goal to reveal the actual content we are interested in. We denote these transformations, which are \textit{not} an integral part of the model's definition, by \textit{normalizations}. 

There is a large amount of freedom in choosing specific normalizations. More specifically, for a function $\varphi:{\mathbb R}^N\to {\mathbb R}^+$, we can define the map $\xi:{\mathbb R}^N\to{\mathbb R}^N$ through
\[\xi[{\boldsymbol y}] = \frac{1}{\varphi({\boldsymbol y})} \,{\boldsymbol y}.\]
Then, evidently, ${\boldsymbol y}\equiv \xi[{\boldsymbol y}]$, implying that
Equation (\ref{eq_ROM_description}) describes identical relative opinion dynamics as
\begin{equation} \label{eq_ROM_description_gamma}
{\boldsymbol y}^{(t+1)} = \xi[ A \,{\boldsymbol y}^{(t)}].\end{equation}
In this paper we typically choose convenient normalizations $\varphi({\boldsymbol y})$, such as keeping the sum of absolute opinion values in $\boldsymbol{y}$ or the Euclidean norm of $\boldsymbol{y}$ constant, thus,
\[\varphi_1({\boldsymbol y}) = \sum_{i=1}^N|y_i|,\hspace{1cm}
\varphi_2({\boldsymbol y}) = \sqrt{\sum_{i=1}^Ny_i^2}.\]

All figures of representations of relative opinions in this paper incorporate a specific choice of normalization $\varphi > 0$, thus  providing the reader a sense of the development of relative opinions as in expression (\ref{eq_relative}). A different choice of $\varphi$ would evidently result in different figures, but the actual content, the ratio of agents' opinions, does not depend on the choice of $\varphi$.

As a consequence of the freedom in choosing normalizations, visualizations of opinion's evolution should be interpreted with care. An agent's opinion (the Outlier) that apparently evolves away from the majority's opinion may actually not be moving at all. Instead, the majority may be shifting its opinion, while the Outlier's opinion remains unchanged. A hypothetical example is a British individual that retains a negative attitude towards homosexual relations since 1983 (see Section \ref{section_motivation}), while the British majority shifts its opinion from negative to positive in the period 1983 to 2010 \citep{Chattoe2014}. From a visualization perspective, the British individual that remains with its negative attitude may have 'moved away' from the majority while not changing its attitude at all.

\vspace{2mm}

\subsection{Example: relative versus absolute opinion framework}
In the following numerical example we show how modeling within a relative opinion framework rather than an absolute framework leads to different conclusions when  ``opinions evolve unboundedly'' in time. More specifically, we show how the relative approach offers a possible---admittedly strongly simplified and speculative---interpretation of the possible dynamics underlying a development like the rise of the \textit{Black Lives Matter} movement against the backdrop of a long historical shift towards increasing consensus on abandoning discriminatory views (represented as opinions in the negative range) on interracial relations in society. 

Consider the opinion dynamics in a population of 100 agents in which there is persistent clustering of opinions within and mutual disagreement between two subgroups of the population, while both subgroups shift over time in the same direction on an underlying opinion dimension\footnote{How exactly this dynamic was generated with our relative opinion model is explained in Appendix \ref{subsect_Fig1Generation}.}. Figure \ref{fig_Change_Abs} shows opinions in an absolute framework. The opinions of both groups exceed any upper bound of a fixed opinion interval within finite time. Substantively, the underlying dynamics can be interpreted as the interaction between a more progressive subgroup in a population and a more conservative one, clustering around two different values on an opinion spectrum. Members of the conservative group (in the model corresponding with numerically lower opinions) are socially influenced by the progressive group (corresponding with numerically higher opinions), so as to gradually shift their opinions towards the progressives. However, the progressive group itself also simultaneously shifts its opinions away from those of its conservative followers, aiming to maintain its relatively more progressive stance in this society. These dynamics described above, viewed from an absolute opinion framework, result in an overall and unbounded increase of the opinion value in both groups over time, while their opinions simultaneously keep diverging from each other. 

In a relative opinion context, only the opinion ratios between agents as in expression (\ref{eq_relative}) are meaningful. To a certain extent, the concept of ``opinions evolve unboundedly'' as in the previous paragraph does not exist from a relative point of view. With identical underlying dynamics, but now in a relative opinion framework the evolution as shown in Figure \ref{fig_Change_Abs} still holds, however, the opinion vectors should now be interpreted as \textit{representative} opinion vectors. Meaning that the actual content lies in the sign of each opinion and the ratios between the opinions. To illustrate the development of the relative opinions we choose a normalization similar to $\varphi_1$. Opinion vectors are normalized such that the sum of the absolute values of the entries  equals the network population size $N=100$; see Figure \ref{fig_Change_Rel} for the corresponding dynamics. The main conclusion is that in relative terms opinions polarize over time to a stable distribution (Figure \ref{fig_Change_Rel}), while in a model based on an absolute interpretation, one of the two subpopulations moves its opinion towards the historical position of the other subpopulation (Figure \ref{fig_Change_Abs}), thus offering a possible interpretation of the dynamics underlying a development like the rise of the \textit{Black Lives Matter} movement. 

\begin{figure}[ht!]
\centering
\begin{subfigure}[t]{0.33\textwidth}
\includegraphics[width=0.95\linewidth]{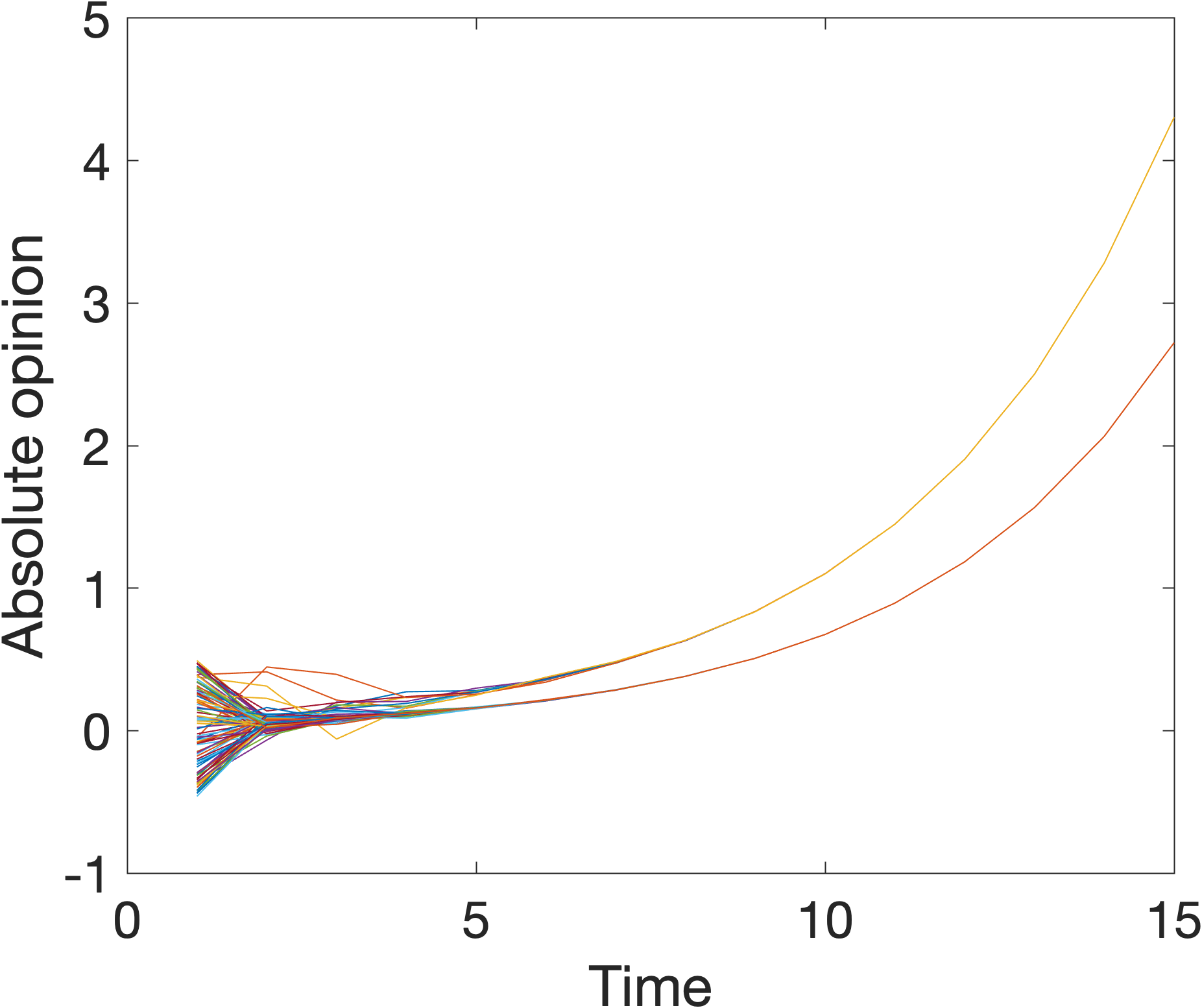} 
\caption{\label{fig_Change_Abs}Model with opinions in an absolute opinion framework}
\end{subfigure}\:\:\:\:
\begin{subfigure}[t]{0.33\textwidth}
\includegraphics[width=0.95\linewidth]{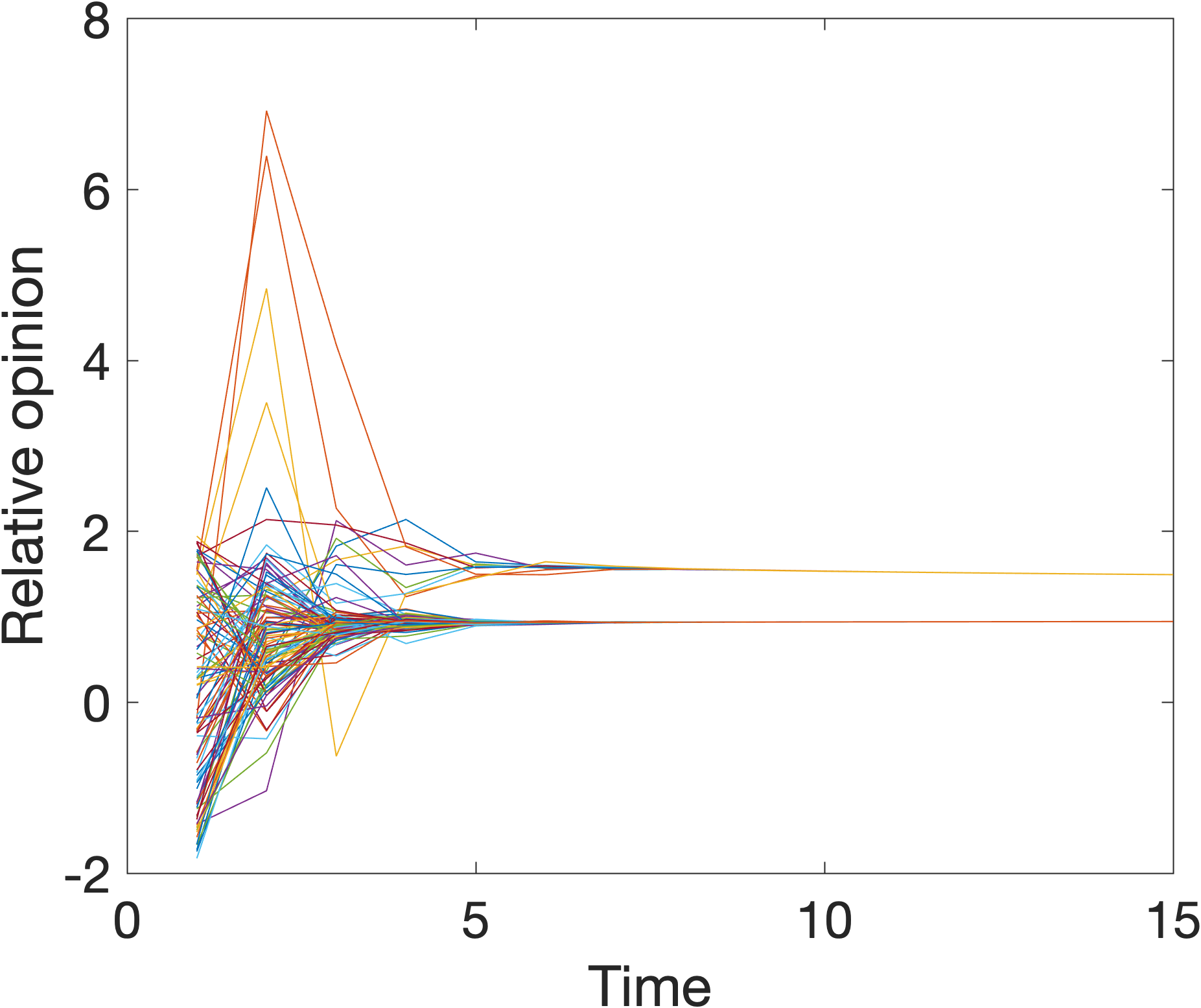}
\caption{\label{fig_Change_Rel}Model with opinions in a relative opinion framework}
\end{subfigure}

\caption{Example of opinion evolution in a population with $100$ agents, in an absolute opinion framework (Figure \ref{fig_Change_Abs}) and in a relative opinion framework (Figure \ref{fig_Change_Rel}) with identical underlying influences. Opinions in a model defined on an absolute opinion framework diverge, while opinions polarize towards a stable difference in a model defined on a relative opinion framework. The representative opinion vectors in Figure \ref{fig_Change_Rel} are scaled such that the sum of their absolute entries equals $100$, meaning that each agent has on average an opinion of $1$ (in absolute value).}
\label{fig_Change}
\end{figure}

\vb

\subsection{Attentiveness and degree of persistence}\label{subsection_attentiveness_persistence}
In this subsection we introduce the terminologies (positive) \textit{attentiveness} and \textit{degree of persistence} which are important parameters in explaining the dynamics of the relative opinion model in the long run, as shown in Section \ref{section_analysis}. Attentiveness is the total ``budget" of susceptibility weights each agent has to distribute among all agents, i.e., the sum of the entries of the agent's row in the matrix $A$. In DeGroot's absolute opinion model each row of $A$  sums to 1, and in the discrete-time Altafini model the sum of the absolute entries of each row sums to 1. In our relative opinion model we allow the sum to be any arbitrary number. 

Here it is borne in mind that a negative row sum is, from an interpretation perspective, implausible. A negative row sum represents an agent that is influenced negatively by the opinion of the entire population on aggregate (including herself). As a consequence, even when the population is in consensus, the agent's opinion changes sign after one time period. This situation is not meaningful theoretically or empirically, hence we will typically think of the row sums as positive numbers ($\sum_j a_{ij} > 0, \, i=1, \ldots,N$), which we will assume from here on. We call the assumption of each agent having positive attentiveness, the \textit{positive attentiveness} assumption. 

An agent's attentiveness reflects the relative importance she assigns to the topic under consideration. The other important agent characteristic is the degree of persistence, describing the degree up to which an agent sticks to her own opinion from the previous time period, reflecting the level of resistance against opinion change. In a sense, the degree of persistence is indicative of the tendency to always take into account a fixed initial opinion value as in \citep{Friedkin1990}. In Section \ref{section_analysis} it is shown how the population's attentiveness and degree of persistence are important parameters in explaining the asymptotic dynamics in the relative opinion model. The formal definitions of the two notions are as follows.

\begin{itemize}
    \item[$\circ$]
Agent $i$'s \textit{attentiveness} is defined as her total susceptibility towards the population, and can be quantified by $\sum_j a_{ij}$. 
\item[$\circ$] Agent $i$'s degree of \textit{persistence} is defined as the level up to which she is resistant towards changing her opinion of the previous time period, or $a_{ii}$. The population's degree of persistence is defined as the sum of the individuals' persistences, i.e., $\sum_j a_{jj} = \Tr{(A)}$.\footnote{$\Tr{(\cdot)}$ refers to the {\it trace function}, which is defined as the sum of the diagonal elements of a square matrix.}
\end{itemize}

\section{Model dynamics} \label{section_analysis}
The aim of this section is to show how the phenomena of polarization, consensus formation, and periodicity may arise in the long run in a relative opinion framework. First, we discuss basic properties of the model at a methodological level. Then, dynamics of the relative opinion model as a result of each agent's attentiveness are explained. Finally, we describe exhaustively (mostly in relation with Theorem \ref{th_k_cap} in the appendix) the asymptotic dynamics of the $N=2$ relative opinion model under the positive attentiveness condition.

\subsection{Basic properties}\label{subsect_basic_char}
As we will argue, no heavy machinery is required for analyses of the relative opinion model; elementary techniques from matrix algebra and Markov chain theory suffice. In particular, the evolution of the opinion vector as described by Equation (\ref{eq_ROM_description}) may be rewritten, under mild regularity conditions (see Appendix \ref{subsec_app_gen_der}), as the sum of the product of its eigenvalues and eigenvectors. As a consequence, Equation (\ref{eq_ROM_description}) takes the form
\begin{equation} \label{eq_evs}
     {\boldsymbol y}^{(t)} = \lambda_1^t b_1 {\boldsymbol v}_1 + \ldots + \lambda^t_N b_N {\boldsymbol v}_N,
\end{equation}
where $\lambda_1, \ldots, \lambda_N$ are the eigenvalues of $A$, ${\boldsymbol v}_1, \ldots, {\boldsymbol v}_N$ its corresponding eigenvectors and $b_1, \ldots, b_N$ coefficients. In the following, the basic characteristics of Equation (\ref{eq_evs}) are informally discussed; for a more formal account we refer to Appendix \ref{subsec_app_gen_der}. 

For the informal discussion, let there be a unique largest eigenvalue $\lambda_1$ (in absolute value) of $A$, with corresponding eigenvector ${\boldsymbol v}_1$. As time progresses, the portion that corresponds to the term of ${\boldsymbol v}_1$ increases, in the sense that eventually ${\boldsymbol y}^{(t)}$ behaves effectively as $\lambda_1^t b_1 {\boldsymbol v}_1$ in the absolute opinion context, entailing that the relative opinions as in Expression (\ref{eq_relative}) converge to a vector proportional to ${\boldsymbol v}_1$. When $\lambda_1$ is positive eventually none of the opinions change sign, whereas when $\lambda_1$ is negative they will keep alternating.  It is also possible  that $\lambda_1$ is complex (but not real valued); then, as explained by Theorem \ref{th_comp_conj}, the relative opinion vector (asymptotically) shows periodic behavior. Summarizing, the long term behavior of the relative opinion vector is determined by the dominant eigenvalue (positive/negative, real/non-real).

Under mild regularity conditions, Equation (\ref{eq_evs}) also applies to the DeGroot and the discrete-time Altafini model\footnote{Here, we refer to the discrete-time Altafini model where the driving matrix is time-independent.}. Therefore, in the study of the development of relative opinions we can build further on results derived from these models, to the extent that conclusions about absolute opinions are allowed to be transferred to relative opinions. In the DeGroot model all entries $a_{ij}$ are positive or zero, with the rows summing to  $1$. 
Hence, after each time step, each agent's opinion has  become either more similar to the opinions of some of the other agents, or has remained completely unaffected (corresponding to a strictly positive or zero influence weight, respectively). For the opinions to reach a consensus, it is therefore sufficient that all agents directly or indirectly influence all the other agents, a condition that is equivalent to the matrix $A$ being irreducible. This mechanism ensures that each agent's opinion moves in the direction of the weighted population mean. However, a complicating factor is that agents can also ``pass around" their opinions periodically without moving in the direction of each other's opinion, a phenomenon referred to as periodicity. Think for example of two agents who fully adopt each other's opinion after each time step. Hence, only if the agents are ``well-connected" (irreducible matrix $A$) and do not just pass opinions around (aperiodicity), then a consensus is reached in the DeGroot model. The translation of the aforementioned to Equation (\ref{eq_evs}) is: there exists a strictly largest real-valued eigenvalue of $A$ and the corresponding eigenvector\footnote{For a non-negative irreducible matrix $A$ with period $h$ and spectral radius $r$ the following holds (Perron-Frobenius): 1.~$r$ is a real-valued eigenvalue with a one-dimensional eigenspace, 2.~$A$ has exactly $h$ complex-valued eigenvalues with absolute value $r$.} is $\boldsymbol{1}_N$ (i.e., the all-ones column vector of dimension $N$). For the case of $N=2$, we will show how consensus formation and conditions such as irreducibility and aperiodicity, as key features of the DeGroot model, reappear in a broader context within the relative opinion model (Appendix \ref{th_k_cap}). The relative opinion model can, to a certain extent, be considered as a generalization of the DeGroot and the discrete-time Altafini model.

\subsection{\textit{Attentiveness} and reaching a consensus}\label{subsect_att} 
An important notion in understanding long term phenomena -- such as consensus formation and polarization -- in the relative opinion model is the level of attentiveness of the population's agents, which we will discuss here. Recall that an agent $i$'s attentiveness $\sum_j a_{ij}$ reflects her total susceptibility towards the opinion of the entire population. An agent with relatively high attentiveness is influenced more by the population's opinion (including herself) than other agents are. From an interpretation perspective, attentiveness provides information about an agent's degree of (net) positive attention towards the opinion topic. Different groups may have a radically different attentiveness regarding specific issues; think of the difference between adolescents and elderly people in their concerns regarding the health consequences of {\sc covid}-19, or the difference between members of minority and majority groups in their concerns regarding the social impact of racial discrimination. 

Two examples are considered here to show how different levels of attentiveness typically result in polarization in the relative opinion model. As in the entire paper, we assume here that all agents have positive attentiveness. In the first example, consider a two agent population where the agents have no interaction with each other, and have different attentivenesses (e.g. $a_{11} > a_{22} > 0$ and $a_{12}=a_{21}=0$). At each time step, the opinion of the agent with the lower attentiveness depreciates relatively to the agent with the higher attentiveness by a factor $\frac{a_{22}}{a_{11}} < 1$. As time progresses, the opinion of the lower-attentiveness agent approaches $0$, and the relative opinion approaches the \textit{representative} opinion vector $(1,0)^\top$\footnote{Observing that the updating matrix is the diagonal matrix $A$ with entries $a_{11}$ and $a_{22}$ with eigenvectors $(1,0)^\top$ and $(0,1)^\top$, the result follows from Equation (\ref{standard_form}) in the appendix.}. This elementary example illustrates how a difference in attentiveness leads to a separation of opinions over time. One could say that it results in polarization, even in a situation in which the agents do not repel each other's opinion, in fact they do not interact at all! 

In the second example, two agents behave essentially identically, but one agent brings twice as much attention to the topic as the other: $a_{1j}=2 \,a_{2j}$, $j = 1,2$. At each time step the opinion of the high-attentiveness agent is twice the opinion of the other agent, so that the relative opinion corresponds to the representative vector $(2,1)^\top$. This illustrates that having different degrees of attentiveness may also lead to polarization when agents do interact.

In fact, consensus formation cannot occur when agents in a population have different levels of attentiveness. This can be easily seen by noting that, under mild regularity conditions, $\boldsymbol{y}^{(t)}$ as in Equation (\ref{eq_evs}) eventually approaches one of the eigenvectors of $A$. Hence, for consensus formation, the all-ones vector $\boldsymbol{1}_N$ must be an eigenvector of $A$ in Equation (\ref{eq_ROM_description}) (see Subsection \ref{subsec_n2} for a definition of consensus formation). This suggests -- to the extent that the model is applicable in real societies -- that consensus will not be reached for topics such as  {\sc covid}-19 measures and impact of racial discrimination as long as the groups have clearly different attention levels, irrespective of the structure of the underlying influence network. As such, one could expect that equal attentiveness could be a condition to reach a consensus by (endogenous) interactions between agents. This would imply for instance that public campaigns for people to stick to {\sc covid}-19 rules would have to start with monitoring and minimizing the different levels of attention across groups in the society, instead of raising attention in general.

The condition of equal levels of attentiveness is ensured in the DeGroot model by assuming the driving matrix $A$ to be a stochastic matrix\footnote{Each row sums to 1.}. A similar assumption holds for the discrete-time Altafini model: here the matrix of which the entries are equal to the absolute value of the entries of $A$, is required to be stochastic, allowing for the possibility of modulus consensus formation \citep{Liu2017}. Since Equation (\ref{eq_ROM_description}) is identical to the model description of the DeGroot and discrete-time Altafini model, but with more general influence weights, the conditions for (modulus) consensus formation applies directly to the relative opinion model (including notions as ``structurally balanced signed digraphs'' \citep{Liu2017}). Under mild regularity conditions, consensus formation in the relative opinion model occurs when the driving matrix $A$ in Equation (\ref{eq_ROM_description}) has an eigenvector that is equivalent to $\boldsymbol{1}_N$ while its associated eigenvalue is positive real-valued and is the unique largest eigenvalue in absolute sense\footnote{Here we note that such uniqueness is not a {\it necessary} condition, see Appendix \ref{subsec_app_gen_der}}, as suggested by Equation (\ref{eq_evs}). Here, we do not attempt to formulate all matrices $A$ with this property, but we emphasize that conditions for consensus formation in the relative opinion model boils down to the matrix algebra question of defining sets of square matrices $A$ with real-valued entries for which the aforementioned holds. The freedom to consider any matrix $A$ in the relative opinion model is an important advantage of the relative opinion model. In Section \ref{sect_groups} we illustrate the convenience of this freedom by showing the possibility of having consensus formation within groups.

Equal attentiveness is a necessary condition for consensus formation in the relative opinion model, but it does not ensure consensus. In particular, in Subsection \ref{subsec_n2} we show that different levels of attentiveness  can also be a driver for asymptotic \textit{periodic} behavior even when the underlying driving matrix $A$ is irreducible and aperiodic. For brevity and readability we call a matrix with equal levels of attentiveness among all agents an \textit{equal attentiveness matrix}. 
We formally say that an $m \times n$ matrix $M$ is a $k$-equal attentiveness matrix if a $k >0$ exists such that $M \boldsymbol{1}_n = k \boldsymbol{1}_m$, with $\boldsymbol{1}_n$ the $n$-dimensional all-ones vector. Going forward we sometimes leave out ``$k$-equal" when it does not cause any ambiguity. This definition also applies to non-square matrices; see Section \ref{sect_groups} for an application.

\subsection{Model behavior for $N=2$}\label{subsec_n2}
This subsection's aim is to exhaustively explain in the $N=2$ relative opinion context how polarization, consensus formation, and periodic behavior arise asymptotically, as a result of the parameters chosen in the updating matrix $A$ appearing in Equation (\ref{eq_ROM_description}). Some  mechanisms align with counterparts in the DeGroot model, whereas others are, to our best knowledge, new within the literature on linear opinion formation models.
We use the following definitions throughout:
\begin{itemize}
    \item[$\circ$] \textit{Consensus formation} is the asymptotic process of opinions moving to within an arbitrarily small \textit{relative} distance of each other\footnote{For any $i,j\in\{1,\ldots, N\}$, $\lim_{t \to \infty} y^{(t)}_i / y^{(t)}_j = 1$ with $y^{(t)}_i$ the $i$-th entry of the opinion vector $\boldsymbol{y}^{(t)}$. Polarization and periodic behavior are mathematically defined in a similar way. An example of consensus formation is the following evolution of representative opinion vector in a two-agent society: $(1,2)^{\top}, (2,3)^{\top}, (3,4)^{\top}, \ldots$. The absolute distance remains 1 while the relative distance evolves as $\frac{1}{2}, \frac{1}{3}, \frac{1}{4}, \ldots$, moving towards an arbitrarily small distance. Formal mathematical definitions of ``convergence'' and ``asymptotic behavior'' are provided in Section \ref{subsubsect_def}.}\label{foot_def_limit}.
    \item[$\circ$] \textit{Polarization} is the asymptotic process of opinions moving to a constant, not arbitrarily small, \textit{relative} distance of each other.
    \item[$\circ$] \textit{Periodic behavior} is the asymptotic process of opinions moving arbitrarily close towards a behavior in which the evolution of relative opinions consistently repeats itself after a fixed time period.
\end{itemize}

Polarization, consensus formation, and periodic behavior are in essence defined as (dynamic) ``shapes of the histogram" of the limiting (for large $t$, that is) opinion distribution. In the situation of consensus formation, this shape converges towards a single vertical line, unlike in the situation of polarization. In the situation of periodic behavior there is no convergence: the shape evolves according to a periodic pattern. Quantitative definitions of polarization in opinion distributions, as observed in surveys, have been proposed  \citep{Bramson2016} and can be applied to characterize stable patterns of polarization in the relative opinions as generated by the relative opinion model, even when there are no stable patterns in absolute opinions.

In the relative opinion model, for the two-agents case the degree of persistence together with the  attentiveness fully determines the asymptotic behavior. Recall that the degree of persistence reflects the degree to which an agent adapts her  opinion  (for agent $i$ quantified by the value of $a_{ii}$ relative to $\sum_{j}a_{ij}$). The degree of persistence of the population is defined similarly ($\sum_i a_{ii} = \Tr{(A)}$). For brevity and readability we introduce the following definitions of three levels of persistence for a two-agents population:

\begin{itemize}
    \item[$\circ$] An agent $i$ is {\it whimsical} when, at each time step, she repels her own opinion ($a_{ii} < 0$). The population is whimsical (on aggregate) when its $\Tr{(A)} < 0$.
    \item[$\circ$] An agent $i$ is {\it open-minded} when, at each time step, she holds on to some of her own opinion while also adopting some of the opinions of the other agents ($a_{ii}\in(0,k_i)$, with $k_i$ the attentiveness of agent $i$, which we assumed to be positive). The population is open-minded (on aggregate) when $\Tr{(A)} \in (0, \sum_i k_i)$.
    \item[$\circ$] An agent $i$ is {\it stubborn} when, at each time step, she strongly holds on to her own opinion while being repelled by the opinions of the other agents ($a_{ii} > k_i$). The population is stubborn (on aggregate) when $\Tr{(A)} > \sum_i k_i$.
\end{itemize}

In combination with attentiveness, these three categories explain exhaustively the qualitative asymptotic behavior of the relative opinion model with two agents under the positive attentiveness assumption. The degree of open-mindedness has been found as an important parameter in related models of opinion change, for example in bounded confidence models \citep{Hegselmann2002, Hegselmann2015}, while stubbornness has been discussed in for example \citep{Flache2004}.  A whimsical agent is similar to a stubborn agent in the sense that their opinion are both repelled from a ``certain'' opinion. However, they are different in the sense that a whimsical agent $i$ tends to change her opinion \textit{towards} the origin ($a_{ii} < 0$), while a stubborn agent $j$ tends to change her opinion \textit{away from} the origin ($a_{jj} > k$). Hence, a whimsical agent has the tendency to change the sign of her opinion, while a stubborn agent has the tendency to stick to the same sign. At the end of this section remarks are made regarding some boundary cases and rates of convergence. It it notable that qualitatively the  results presented in this section extend to populations of arbitrary large $N$ that consist of two groups, as will be extensively discussed in Section \ref{sect_groups}.

In the discussion below we do not include the trivial case that $A$ represents two agents with equal attentiveness and no interaction. In this trivial case, where $A$ effectively corresponds to the identity matrix, clearly the opinion vector remains constant over time; it corresponds to the two-agents DeGroot model with the updating matrix $A$ being reducible. The results as discussed in the following subsections are all under the positive attentiveness assumptions; we write ``relative opinion model'' for the relative opinion model under the positive attentiveness assumption. All results are substantiated in the appendix, in particular by Theorem \ref{th_k_cap}.

\subsubsection{Consensus formation}\label{subsubsect_cons}
\begin{figure}[htbp]
\centering
\includegraphics[width=0.33\linewidth]{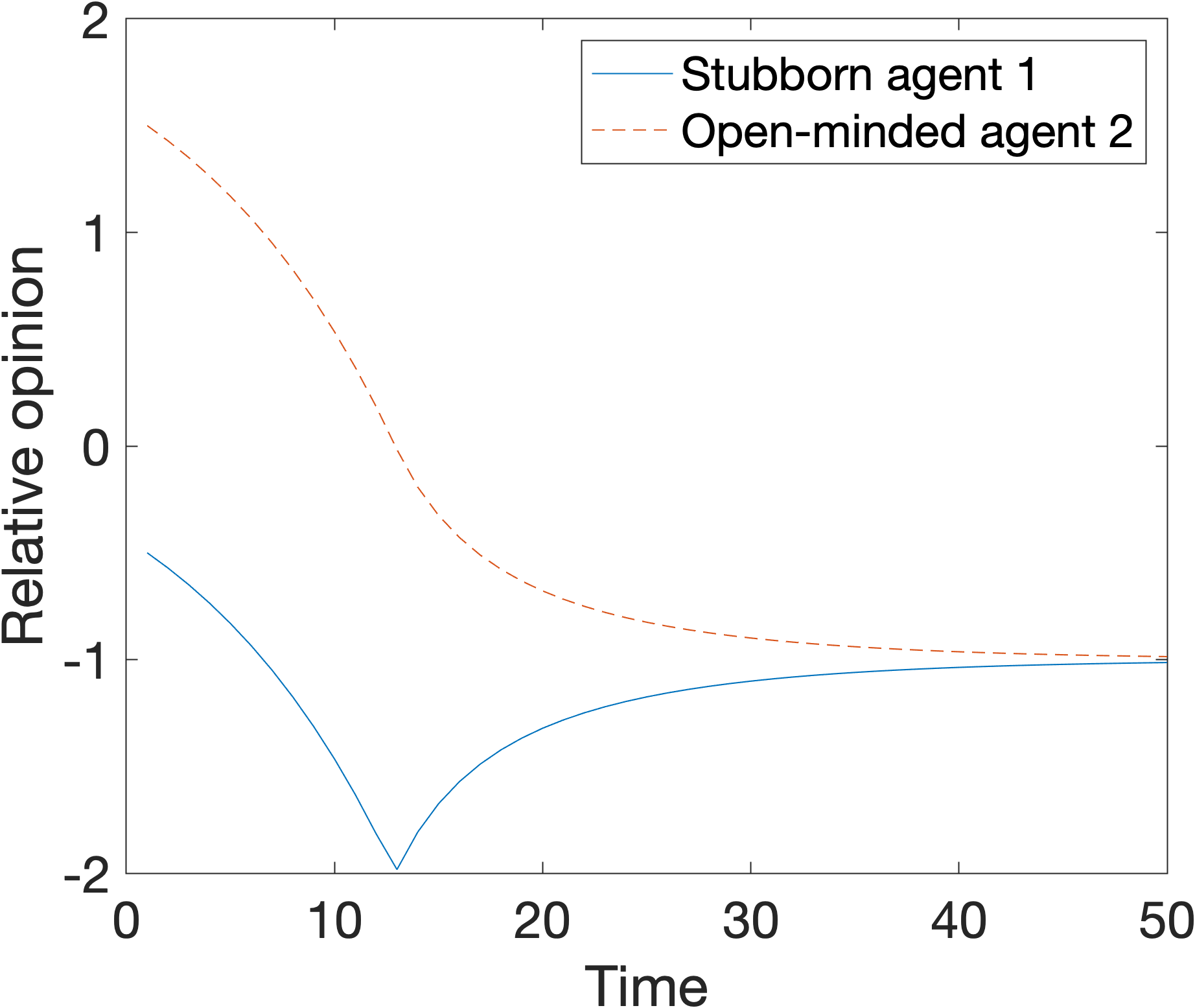} 
\caption{\label{fig:image21}Consensus formation in a population of two agents with equal attentiveness. The opinion vectors are scaled such that the sum of their absolute values equals $2$. The population is open-minded, still it consists of an open-minded and stubborn agent.}

\end{figure}

The relative opinion model for two agents leads to consensus formation if two conditions are met. The first of these conditions is equal attentiveness of all agents, as discussed in Subsection \ref{subsect_att}. The necessary additional condition for consensus formation is open-mindedness: the agents must \textit{on aggregate} be open-minded. Notably, it is not necessary for both agents to be open-minded.
These two conditions together are similar to the conditions under which opinions in the two-agent DeGroot model reach a consensus. There is a crucial difference, though:  in the DeGroot model both agents must be open-minded, while in the relative opinion model consensus can even be reached when one of the two agents is not; see Figure \ref{fig:image21}.

\subsubsection{Polarization}
\begin{figure}[htbp]
\begin{subfigure}[t]{0.32\textwidth}
\includegraphics[width=0.95\linewidth]{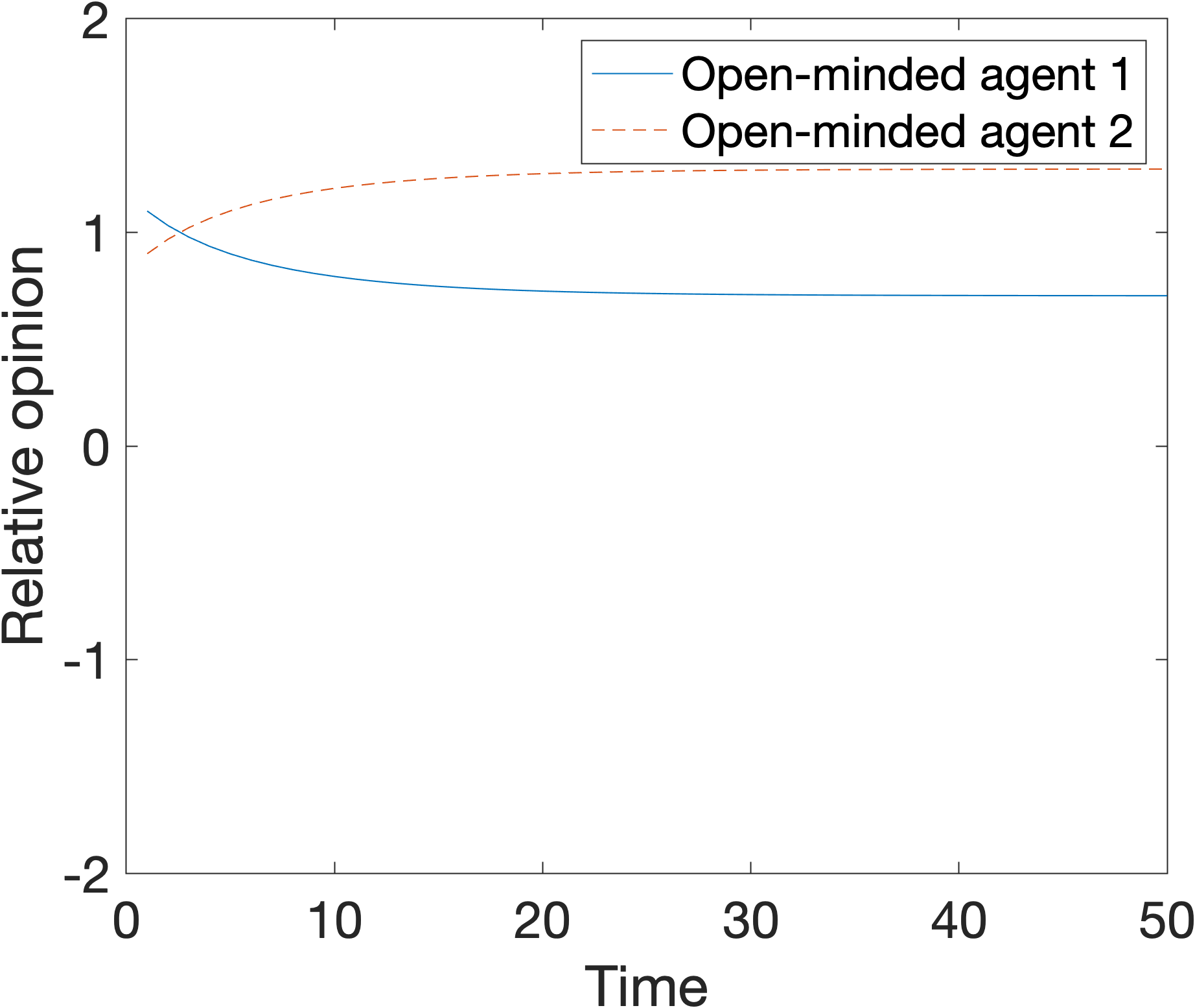} 
\caption{Type 1: unequal attentiveness}
\label{fig_pol_uneq}
\end{subfigure}\:\:
\begin{subfigure}[t]{0.32\textwidth}
\includegraphics[width=0.95\linewidth]{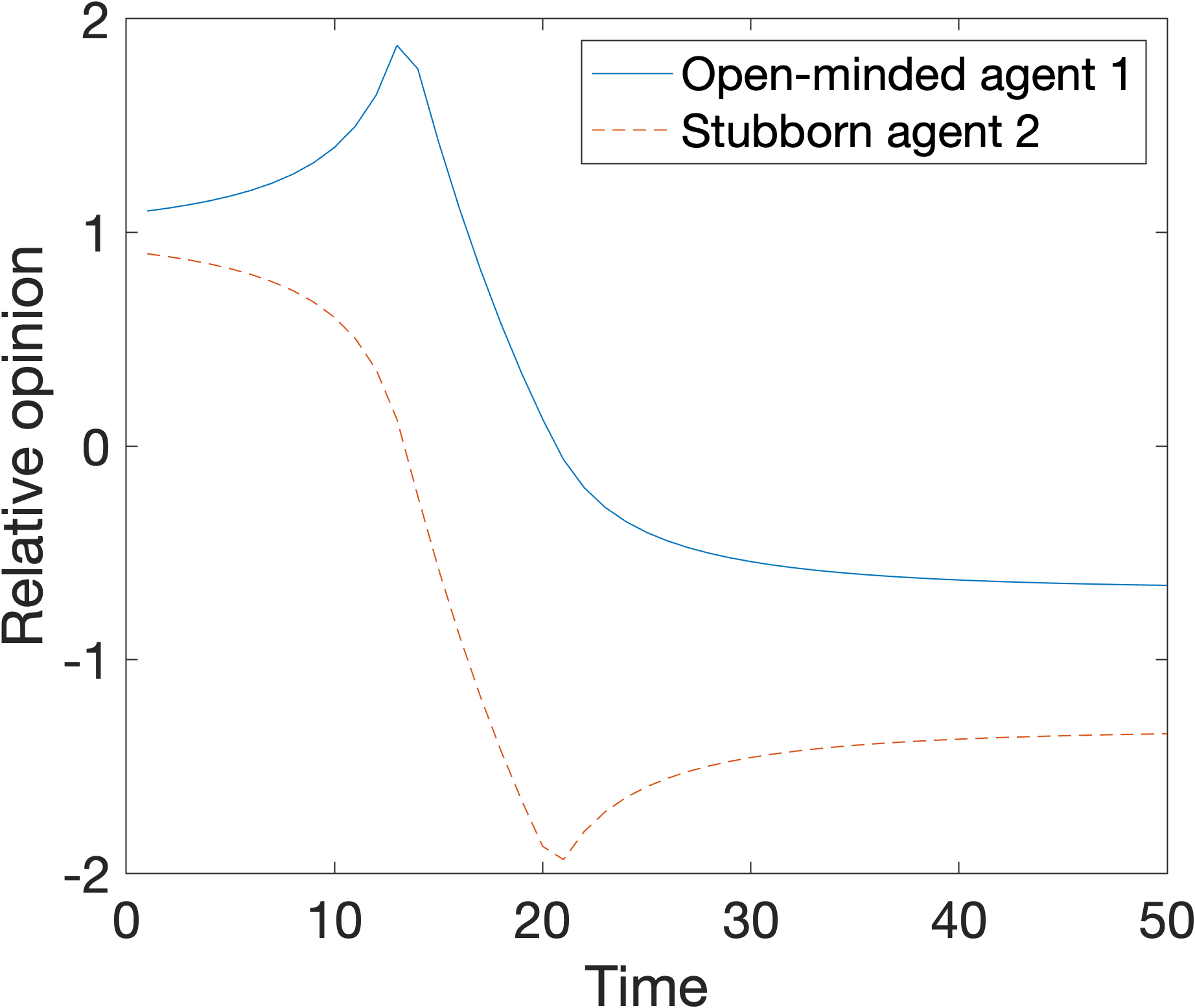}
\caption{Type 2: equal attentiveness and stubborn population with one stubborn agent}
\label{fig_pol_eq_op_st}
\end{subfigure}\:\:
\begin{subfigure}[t]{0.32\textwidth}
\includegraphics[width=0.95\linewidth]{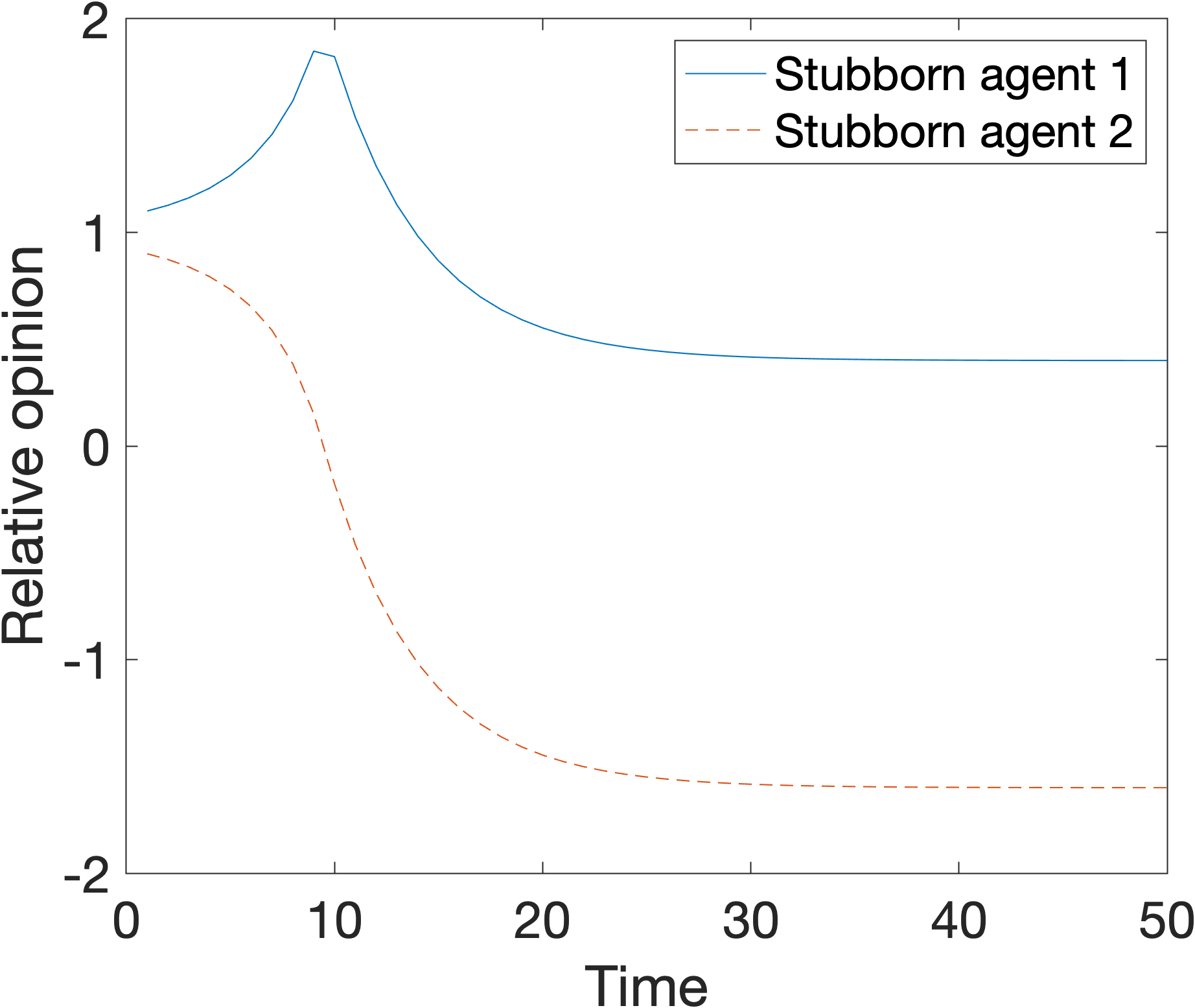}
\caption{Type 3: equal attentiveness and stubborn population with two stubborn agents}
\label{fig_pol_eq_st_st}
\end{subfigure}

\caption{Polarization in a population of two agents. The opinion vectors are scaled such that the sum of their absolute values equals $2$. Polarization in Figure \ref{fig_pol_uneq} arises as a result of different attentiveness between the agents. Polarization in Figure \ref{fig_pol_eq_op_st} arises under equal attentiveness and a stubborn population; here one agent is stubborn and one is open-minded. Polarization in Figure \ref{fig_pol_eq_st_st} arises under equal attentiveness and a stubborn population; here both agents are stubborn.}
\label{fig_pol}
\end{figure}

The relative opinion model for two agents allows for three types of polarization, as illustrated by Figure \ref{fig_pol}. The first type (Figure \ref{fig_pol_uneq}) results from agents having unequal degrees of attentiveness, as explained in Subsection \ref{subsect_att}. We remark that in certain circumstances, unequal attentiveness can also result in periodic asymptotic behavior, as depicted in Figure \ref{fig_per_uneq_op_st}; we get back to this in Subsection \ref{subsect_per}.

The second and third types of polarization (Figures \ref{fig_pol_eq_op_st} and \ref{fig_pol_eq_st_st}) arise as a result of equal attentiveness and a stubborn population. In case the stubborn population  consists of an open-minded and stubborn agent, we have polarization of type 2. Observe from Figure \ref{fig_pol_eq_op_st} that in this case the signs of the limiting relative opinions remain equal. To have polarization of type 3, i.e., limiting relative opinion of different signs, it is required that both agents are stubborn; see Figure \ref{fig_pol_eq_st_st}. As in Figure \ref{fig_Change_Abs}, the parameters applied in these examples would result in absolute opinions to grow beyond any bound, but within a relative context the opinions stabilize.

Type 1 polarization directly relates to earlier models of opinion dynamics addressing persistent opinion variation as a consequence of strategically changing salience of an issue for agents \citep{Flache2004, Stokman1995}. Polarization of types 2 and 3, being  a consequence of negative social influences, has been studied before in opinion formation models \citep{Flache2011b, Baldassarri2007, Mark2003}. One typical outcome of these models is two clusters at maximal distance within a confined absolute opinion space, but also fragmentation into multiple opinion clusters can occur \citep{Mas2014}. Polarization in the relative opinion model is more general in the sense that it also covers persistently shifting opinions of both groups in absolute terms, but converging to a constant distance in the relative opinion context.

\subsubsection{Periodicity}\label{subsect_per}
\begin{figure}[ht]
\begin{subfigure}[t]{0.32\textwidth}
\includegraphics[width=0.95\linewidth]{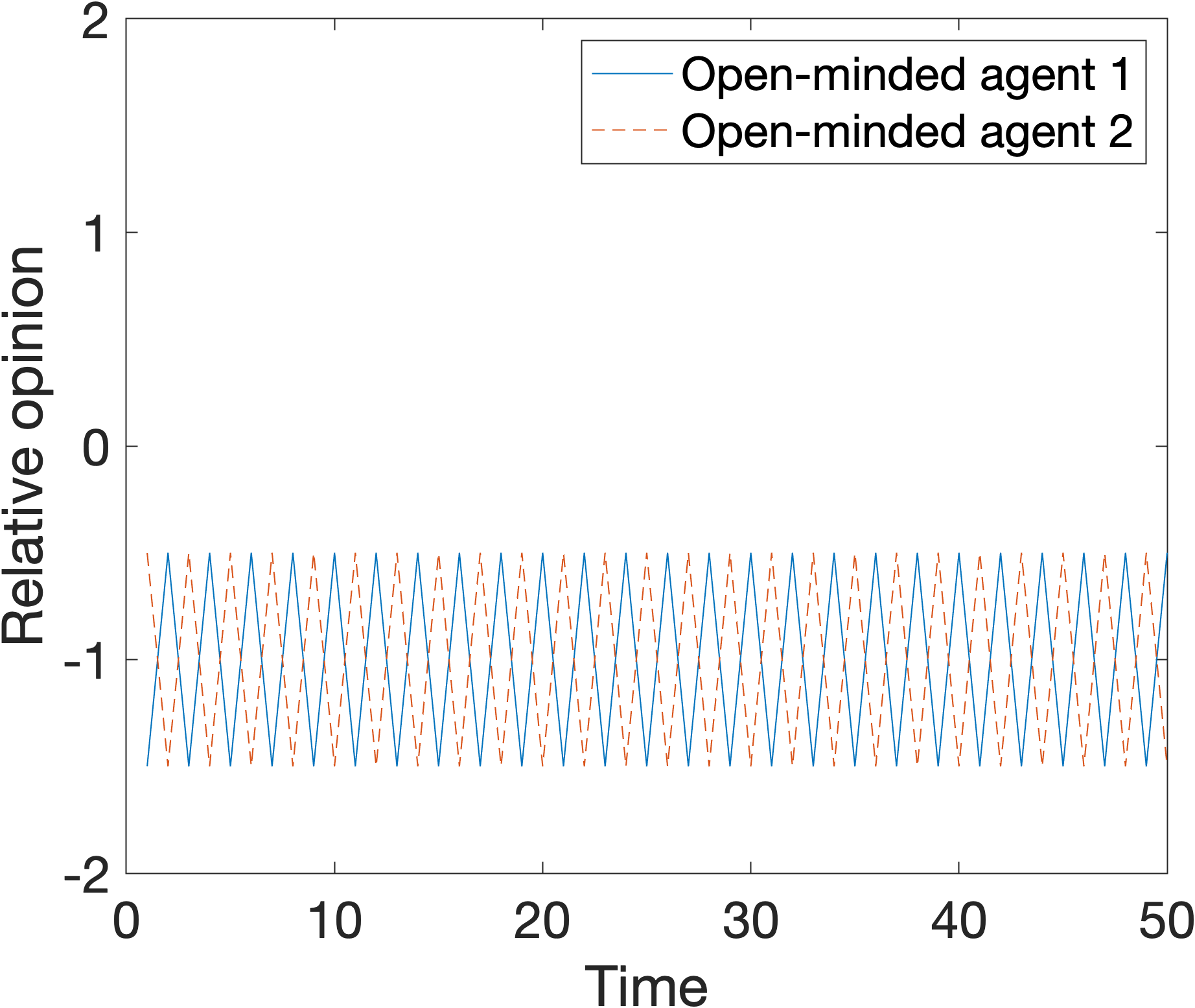} 
\caption{Type 1: equal attentiveness and artificially passing around opinions}
\label{fig_per_pass}
\end{subfigure}\:\:
\begin{subfigure}[t]{0.32\textwidth}
\includegraphics[width=0.95\linewidth]{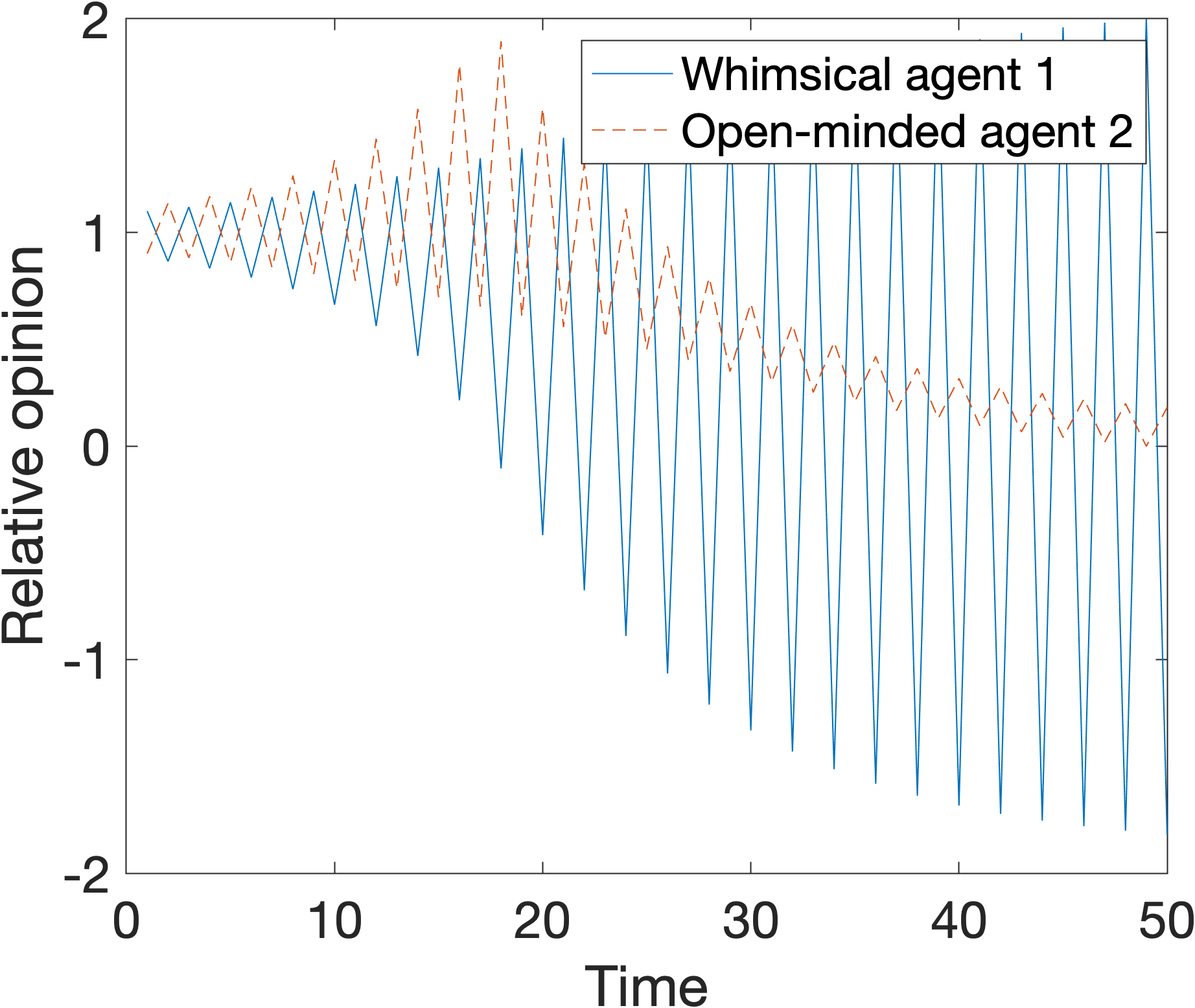}
\caption{Type 2: equal attentiveness and whimsical population}
\label{fig_per_eq_wh_op}
\end{subfigure}\:\:
\begin{subfigure}[t]{0.32\textwidth}
\includegraphics[width=0.95\linewidth]{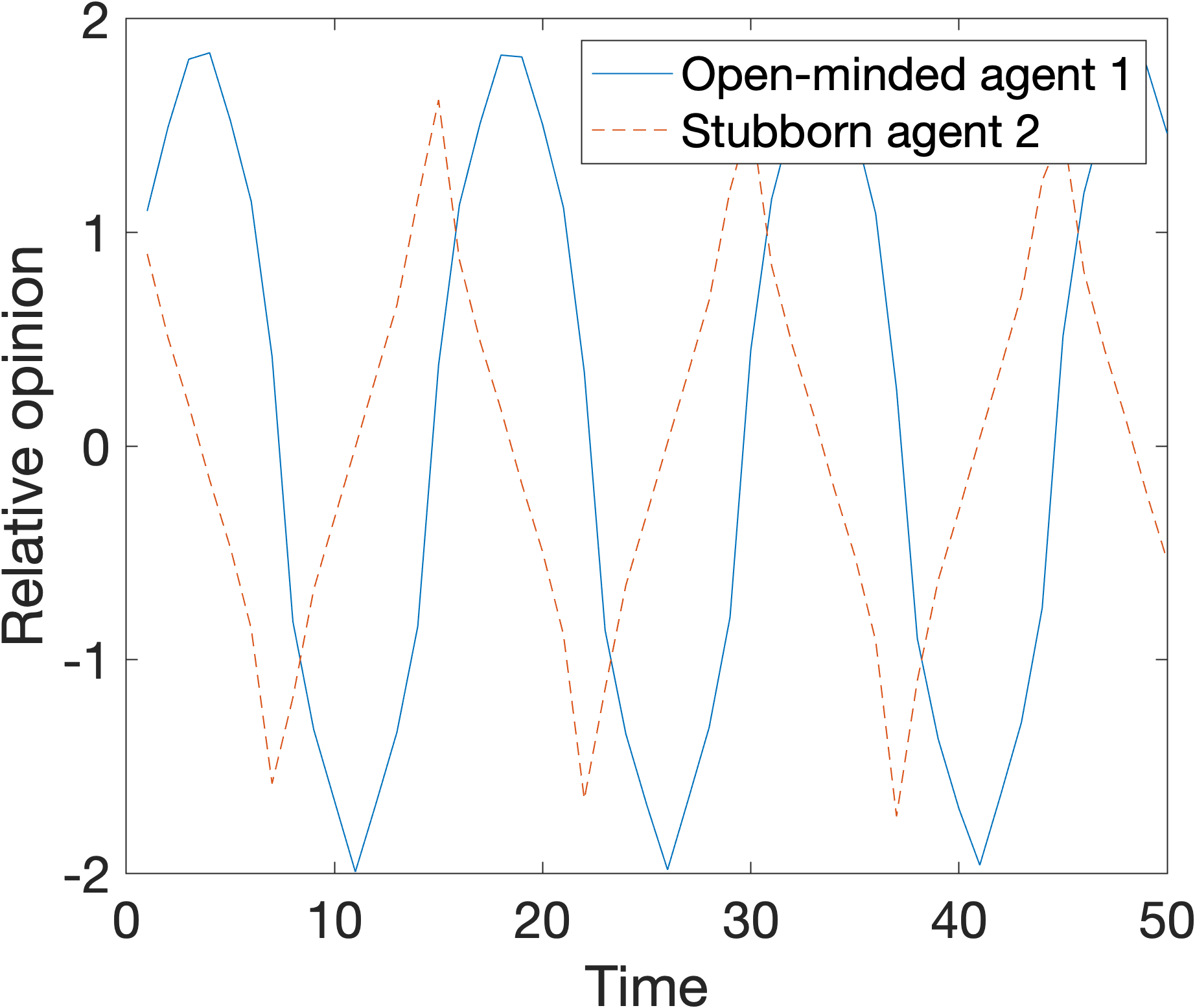}
\caption{Type 3: unequal attentiveness combined with "strong" stubborn and open-minded interaction}
\label{fig_per_uneq_op_st}
\end{subfigure}

\caption{Periodic behavior in a population of two agents. The opinion vectors are scaled such that the sum of their absolute values equals $2$.  Periodic behavior in Figure \ref{fig_per_pass} arises as a result of equal attentiveness and open-minded agents who completely adopt the opinion of the other agent. Periodic behavior in Figure \ref{fig_per_eq_wh_op} arises as a result of equal attentiveness and a whimsical population. Periodic behavior in Figure \ref{fig_per_uneq_op_st} appears under unequal attentiveness combined with ``strong" stubborn and open-minded interaction, i.e., $(a_{11} - a_{22})^2 < -4a_{12}a_{21}$.}
\label{fig_per}
\end{figure}

Asymptotic periodic behavior of the relative opinion model of two agents appears as a result of three different types of mechanism. In the first type, displayed in Figure \ref{fig_per_pass}, each agent adopts the other agents' beliefs completely, as if each of them passes on the others' opinions at each time period. This type of periodicity is also covered by the DeGroot model.

The second type of periodic behavior appears when the population is whimsical and all agents have equal attentiveness; see Figure \ref{fig_per_eq_wh_op}. Here, the opinion of the whimsical agent follows an alternating pattern, while the open-minded agent follows the opinion of the whimsical agent, thus displaying an alternating pattern as well. One could argue that a population that is whimsical may be considered as unrealistic. Still, it may represent a hypothetical population that predominantly consists of ``pioneers'' who have an innate urge to change the sign (from positive to negative and vice versa) of their opinions.

The third type of periodic behavior, and the only type that allows for periods larger than two time steps, arises as a result of unequal attentiveness and a ``strong" enough interaction between the stubborn and open-minded agent. The situation of the relative opinion evolution, as illustrated in Figure \ref{fig_per_uneq_op_st}, can be interpreted as a stubborn agent that tries to move away from the open-minded agent's opinion, first by evolving to an opinion with a different sign. Since the open-minded agent follows the stubborn agent, after a while the signs of both agents will be equal; the stubborn agent then repeats her behavior by moving her opinion to the other side of the opinion spectrum, thus leading to periodic behavior. 

A counterforce to type 3 asymptotic periodic behavior is the difference in degree of persistence between the agents. As discussed in Subsection \ref{subsect_att}, in a population with no interaction, different degrees of persistence result in decay of the low-persistence agent's opinion to a neutral position. When the agents' degrees of persistence ``sufficiently'' differ, this force dampens the forces involved in the type 3 periodicity mechanism, producing consensus formation when the population is open-minded and has equal attentiveness (see Subsection \ref{subsubsect_cons}) and polarization otherwise. Mathematically, periodicity of type 3 arises when $(a_{11} - a_{22})^2 < -4a_{12}a_{21}$, (Proposition \ref{prop_comp_22})  where the left-hand side represents the difference in degree of persistence and the right-hand side the degree of the forces that cause type 3 periodic behavior.

Under the condition of equal attentiveness (i.e., $a_{11} + a_{12} = a_{21} + a_{22} = k$), the difference in degree of persistence (i.e., $(a_{11} - a_{22})^2$) increases with an increase in the level of interaction  between the stubborn and open-minded agents (i.e., $4a_{12}a_{21}$). Adding the forces resulting from unequal attentiveness is therefore necessary in order to produce type 3 periodic behavior. 

Asymptotic periodic behavior of type 3 seems to reflect the dynamics between a typical pioneer and the masses which tend to follow the pioneer's ideas. In this context, one could think of non-fashionistas copying the ideas of fashionistas in the fashion industry, or regular social media users who follow the actions of influencers.  Observe that the pioneer in the type 3 example changes her opinion in the opposite direction, relative to those of  the \textit{other} agents, while the pioneer in the type 2 example changes her opinion opposite to her \textit{own} opinion.

Investigation of patterns of persistent instability in opinion formations has been performed before, for example in  \citep{Flache2004} or  \citep{Strang2001}. However, to our best knowledge, it has never been demonstrated before in a linear model with a ``well-connected" and ``aperiodic'' population.
As in Figure \ref{fig_Change_Abs}, the applied influence weights in producing Figure \ref{fig_per} leads to opinions growing unboundedly in time considered from an absolute opinion interpretation; within the relative context the opinion development stabilizes to a periodic pattern.

\subsubsection{Remark: boundary cases and rate of convergence}
The description of the asymptotic behavior of the relative opinion model for two agents under the positive attentiveness assumption in the previous subsections is ``near-exhaustive'': what remains open are the boundary cases where, under equal attentiveness, the degree of persistence of the population is exactly between whimsical and open-minded, and between open-minded and stubborn. As shown by Theorem \ref{th_k_cap},  in the appendix, in these boundary cases our main findings essentially carry over.

The rate at which the relative opinion vector starts to reveal consensus formation, polarization or periodic behavior is typically exponential, as can be seen from Equation (\ref{eq_evs}). Under the assumption of equal attentiveness, the rate is relatively low when the degree of persistence of the population is ``close" to the boundary values. This is explained by observing that the forces of being whimsical, open-minded, and stubborn are weaker around the boundary values of degrees of persistence. When the degree of persistence is exactly between open-minded and stubborn, the rate becomes $O(t^{-1})$, whereas, remarkably, when the degree is exactly between whimsical and open-minded, ``convergence" is immediate. 

\section{Analysis of dynamics in a two-group population} \label{sect_groups}
In this section we extend the results of the two-agent relative opinion model, as was discussed in Subsection \ref{subsec_n2}, to a two-\textit{group} population with arbitrarily many agents. The theory of this section serves as a basis for producing opinion dynamics as described by Figure~\ref{fig_Change}. Our results on groups illustrate that the relative opinion model still allows relatively explicit mathematical analysis for arbitrarily large population sizes and relevant structures of the updating matrix $A$. This section also provides techniques to produce more complex asymptotic dynamics than the ones in the two agent population. Additionally, it contains a hypothesis on when the individual groups' asymptotic behavior carries over to the entire population.

\subsection{Definition of groups}\label{subsect_def_groups}
The terminology ``groups of individuals" is used in a variety of contexts. Examples are groups of individuals who are in favor or against the Black Lives Matter movement, or individuals who believe or do not believe in the safety of vaccination. 
One of the main conceptualizations of polarization in the literature is that disagreement between groups occurs together with agreement within groups \citep{Flache2017, Bramson2016, Esteban1994}. This seems to coincide with patterns of opinion distribution often encountered regarding contentious issues: for example, during the protests of the Black Lives Matter movement  the opinions of supporters more or less reached a consensus, while the opinions of the opponents did so as well. This motivates why  the definition of groups should at least cover consensus formation {\it within} the groups simultaneously with polarization {\it between} the groups.

As shown in Subsection \ref{subsect_att}, agents \textit{must} have an equal degree of attention to allow for consensus formation. In the relative opinion framework, it is therefore necessary to require that agents belonging to the same group have equal attentiveness towards their own group. As shown in the appendix (\ref{th_22ev}), an equal attentiveness towards the agents of the other group as well may lead to polarization between groups, while consensus formation occurs within the groups. Therefore, groups in this section are characterized by the agents' degree of attention.

Throughout this section, the following definition will be used. In a two group setting, a group is  a collection of agents that have equal attentiveness to the agents in their own group and have equal attentiveness to the agents of the other group. The attentiveness of the group's agents to their own group may differ from the attentiveness to the agent of the other group.
Evidently, a population with two groups can be rearranged such that the updating matrix consists of $4$ blocks of equal attentiveness matrices:
\begin{equation}\label{eq_2block_intro}
    A = 
    \begin{pmatrix}
        A_{11} & A_{12} \\
        A_{21} & A_{22}
    \end{pmatrix}.
\end{equation}
Here, $A_{11}$, $A_{22}$ are square $c_{11}$- and $c_{22}$-equal attentiveness matrices, and $A_{12}$ and $A_{21}$ are (not necessarily square) $c_{12}$- and $c_{21}$-equal attentiveness matrices. The coefficients $c_{11}, c_{12}, c_{21}, c_{22} \in {\mathbb R}$ are called \textit{attentiveness degrees}.
As the matrices $A_{11}$ and $A_{22}$ contain the set of weights that describe the influence between agents within their group, we call them \textit{intragroup matrices}. By the same token we call $A_{12}$ and $A_{21}$ \textit{intergroup matrices}. 

Relying on the definition of equal attentiveness, it is possible to decompose each intra- and intergroup matrix into its attentiveness degree and $1$-equal attentiveness matrix:
\begin{equation}\label{eq_2block_outgroup}
    A = 
    \begin{pmatrix}
        A_{11} & A_{12} \\
        A_{21} & A_{22}
    \end{pmatrix} =
    \begin{pmatrix}
        c_{11} P_{11} & c_{12} P_{12} \\
        c_{21} P_{21} & c_{22} P_{22}
    \end{pmatrix},
\end{equation}
with $P_{ij}$ $1$-equal attentiveness matrices, $i,j\in\{1,2\}$. 
Equation (\ref{eq_2block_outgroup}) reveals that within each group the dynamics are governed by $P_{11}$ and $P_{22}$, while $P_{12}$ and $P_{21}$ determine the dynamics between the groups, within the structure imposed by the matrix of attentiveness degrees:
\begin{equation}\label{eq_group_lvl}
    C = 
    \begin{pmatrix}
        c_{11} & c_{12} \\
        c_{21} & c_{22}
    \end{pmatrix}.
\end{equation}

A comparable group setting is considered analytically in for example  \citep{EGER2016} and through simulations in  \citep{Amblard2004}. An example where this group framework may be suitable relates to the opinion formation dynamics of a population consisting of fashionistas and non-fashionistas. The group of fashionistas (say, group 1) typically has a smaller population size than the group of non-fashionistas (group 2). Roughly speaking, fashionistas have a relatively high degree of attention to fashion-related issues (``What shape should glasses have?'') than non-fashionistas, in our framework reflected by $c_{11} + c_{12} \neq c_{21} + c_{22}$. Also, fashionistas typically look for a style that differs from the mainstream one ($c_{12} < 0$), while non-fashionistas have the tendency to adopt the new styles of the fashionistas ($c_{21} > 0$). The matrix of attentiveness degrees $C$ then has unequal values of attentiveness and, with ``sufficiently strong" stubborn and open-minded interaction, the result we found for the two agent population in Subsection \ref{subsect_per} suggests that the population may show asymptotic periodic behavior. These dynamics could help interpret, for example, the recurrence of different colors and styles in the fashion industry over the past decades. 

\begin{figure}[t]
    \centering
    \includegraphics[width=13cm]{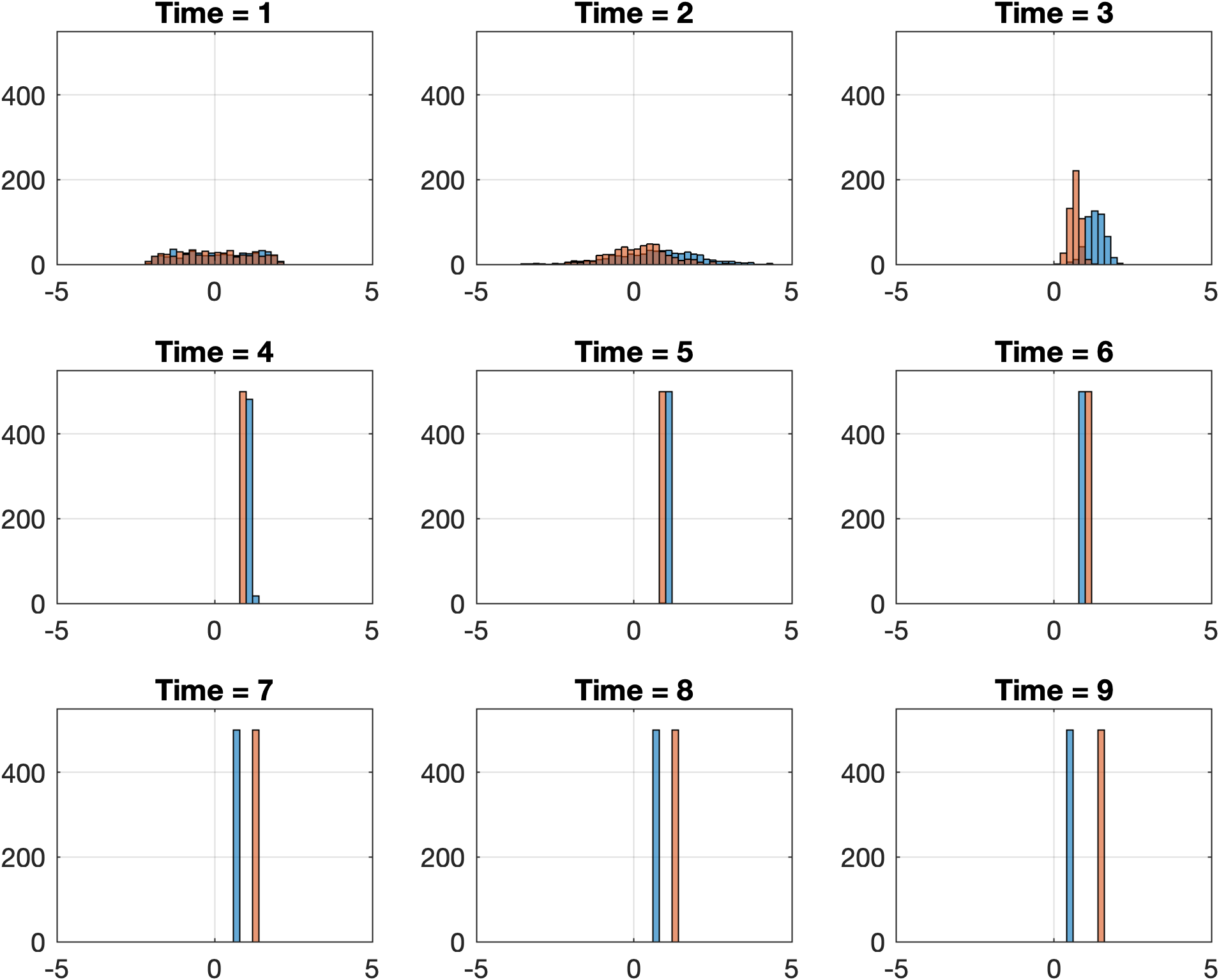} 
    \caption{Histogram of opinions of two ``well-connected'' groups at different moments in time. The blue bars depict group 1 of size 500 representing a group that repels the opinion of the other group (e.g.\ fashionistas), the orange bars depict group 2 of size 500 representing a group that follows the opinion of the other group (e.g.\ non-fashionistas). Opinion vectors are scaled such that the sum of its absolute entries equals 1000. The entries of the intragroup $1$-equal attentiveness matrices $P_{11}$, $P_{22}$ and intergroup $1$-equal attentiveness matrices $P_{12}$, $P_{21}$ are chosen uniformly between $-0.38$ and $0.62$, i.e., with a slight positive drift. Each row is then scaled to satisfy the $1$-equal attentiveness condition. The matrix of attentiveness degrees $C$ has entries $c_{11} = 1.5$, $c_{12} = -0.3$, $c_{21} = 0.5$, and $c_{22} = 1$.} 
    \label{fig_group_500_500_fashion}
\end{figure}

Figure \ref{fig_group_500_500_fashion} shows the result of a simulation of opinion formation dynamics where agents are divided into two groups according to the agents' attentiveness, but still are well-connected. The blue bars represent the fashionistas and the orange bars the non-fashionistas.  After 1 unit of time, the opinions are more or less uniformly distributed around the origin, two time units later agents within the groups are showing signs of consensus formation, and again two time units later the agents within the groups have reached a consensus. From 6 time units onward, the stubborn group moves toward the origin because it wants to move away from the opinion of the open-minded agents. Depending on the exact entries in the matrix $A$, this can play out as  complex cyclic dynamics in the bounded relative opinion space, in which fashionistas move away from non-fashionistas, while non-fashionistas follow the fashionistas. In absolute opinion space, however, such dynamics could eventually move away from any finite interval.

\subsection{Transfer of group's asymptotic behavior}\label{subsect_group_transfer}
Even in the restricted setting of two groups, it is complicated to analytically investigate the asymptotic opinion formation dynamics of a population with arbitrarily many agents. One way to still gain insight into the asymptotic dynamics of a two-group population is by decomposing the intragroup effect from intergroup effects. 
Indeed, we  express the eigenvalues of  the influence matrix $A$ of the entire population, in terms of (i)~the eigenvalues of the intragroup matrices $A_{11}$ and $A_{22}$ as if they were stand-alone populations, and (ii)~the influence between the groups as described by the attentiveness degrees in $C$.

As explained in Subsection \ref{subsect_basic_char}, the largest eigenvalue of $A$ (in absolute terms, that is) in Equation (\ref{eq_ROM_description}) determines the asymptotic opinion formation behavior. Thus, being able to express the eigenvalues of the entire population in terms of the eigenvalues of the intragroup matrices implies that the asymptotic behavior of the entire population is fully determined by the asymptotic behavior of its groups.

An example is a company that is divided into a group of leaders and a group of employees, together constituting the population of agents. In the hypothetical case that employees do not have any preference to follow a specific leader (in that they tend to follow the opinion of the group of leaders as a whole), the employees eventually will have an opinion that aligns with that of the leaders. However, it is possible that within the group of leaders, some are stubborn and argue regularly, which results in polarization within the group of leaders. As a reaction, the employees will also polarize since they follow the opinions of the leaders. And, since the population considered consists of leaders and employees only, the company becomes polarized. In this example, one may argue that the dynamics within the group of leaders is dominant for the asymptotic opinion dynamics. From a mathematical perspective, one would expect that the largest eigenvalue (in absolute terms) of the influences between leaders plays a crucial role in the largest eigenvalue of the entire population.

Another example concerns the interaction between fashionistas and non-fashionistas, as illustrated in Figure \ref{fig_group_500_500_fashion}. Group members tend to reach a consensus with fellow group members, while fashionistas have the inclination to move away from the opinion of non-fashionistas, a behavior also called ``individualization" in Mäs et al. 2010 \citep{Mas2010}. The resulting asymptotic periodic behavior of the entire population follows from the interaction between groups as described by the attentiveness degrees in $C$. Mathematically, it is expected that in this case the largest eigenvalue of $A$  is predominantly the result of the  largest eigenvalue of $C$.

It follows from Theorem \ref{th_22ev} in the appendix that in the special case where fellow group members are influenced identically by members of the other group at each time period, the eigenvalues of $A$ can be expressed in terms of the eigenvalues of $P_{11}$, $P_{22}$ and $C$. Mathematically, the condition of being influenced identically by members of the other group corresponds to $P_{12}$ \textit{or} $P_{21}$ being matrices with identical columns. One can interpret this as follows. Each fellow member of one group is not in individual contact with members of the other group. Rather, the influence from non-group members is channeled through a central source such as a spokesperson or mass media. 

Arguably, in real life typically the condition of identical influences by members of the other group does not hold exactly. Still it provides insights into how group behavior may propagate to the entire population. The sensitivity analysis presented in Appendix \ref{subsect_SensA}, where the identical columns of $A_{12}$ are increasingly perturbed by a noise factor of up to 120 percent of its attentiveness degree, shows that the eigenvalues are still transferred from $P_{11}$, $P_{22}$ and $C$. Another sensitivity analysis on the effects of the equal attentiveness assumption on $A_{11}$, $A_{12}$, $A_{21}$ and $A_{22}$ has been performed, where the sums of rows differed up to 3 percent of its attentiveness degree. Also in this case the transfer of eigenvalues was robust. Thus, even in cases in which the identical influences condition is not exactly met, our model still provides an accurate reflection of the asymptotic behavior.

We also explored the case where agents of different groups do not interact directly, but they follow media which in turn interact with media followed by the other group. Simulations show that the intragroup matrices of $A^t$ (i.e., $A$ multiplied $t-1$ times with itself), for time $t$ larger or equal than 3 effectively satisfy the condition of identical influence by members of the other group. Only after more than 3 time periods the media starts to transfer the opinion of its group. In this case it is possible to construct a population that polarizes into two opinions as a result of the interaction between media sources. In fact, when media are the sole source of influence of agents in a group by another group, any type of asymptotic dynamics is possible, depending on the type of interaction between the media sources (as described by $C$). The analytical result of Theorem \ref{th_22ev} effectively reveals that more influence by media results in a more pronounced transfer of the asymptotic dynamics from the groups to the entire population.

For symmetry reasons we expected that any condition on the transfer of eigenvalues of the groups to the entire population is bilateral, which implies in this case that identical influence should go both ways. It is therefore rather unanticipated that unilateral identical influence of group members (which is a condition on \textit{either} $P_{12}$ \textit{or} $P_{21}$) is already sufficient. This suggests the hypothesis that in settings where members of one of two interacting groups perceive influences from those of the other group as if they come from a unitary source (each member of the other group being a unitary source), for example through stereotyping or media influence, this forces the other group into a pattern where influences from members of the former group are likewise responded to as if they come from a unitary actor. A tentative interpretation of this could be that unilateral stereotyping of an outgroup in a two-group setting results in behavior similar to what can be expected under mutual stereotyping.

The analytical result (Theorem 
\ref{th_22ev}) on the ``transferability'' of asymptotic dynamics shows that asymptotic dynamics of a two-group population may be predominantly caused by either the interaction within the groups or the interaction between the groups. Evidently, in models with a more involved structure (rather than two groups), one could encounter intrinsically more complex behavior.

\section{Discussion}
An outstanding research problem in the field of opinion formation processes concerns finding an elementary linear framework that is capable of reproducing the main large-scale phenomena of opinion dynamics (viz.\ polarization, consensus formation, and periodicity) in settings in which the population is ``well-connected" and ``aperiodic''. In this paper we contributed to addressing this challenge: we have developed a relative opinion model that exhibits all desired large-scale phenomena while still working with a linear update rule. 

In addition, the relative opinion model allows to identify stable patterns in opinion formation that occur over a longer time scale, even when, seen from a particular point in time, opinions shift outside a fixed range (as represented by scales like those typically used in opinion surveys). Such longer term shifts may occur because, for example, societal norms on certain issues change over time so that opinion distributions keep shifting to different sections of the same underlying opinion dimension. Models representing only absolute opinions can not readily describe patterns that are stable despite those shifts. 

The main technical reason for the richness of the relative opinion model's dynamics is the inclusion of the possibility of a repulsive force between agents, while remaining in a linear framework. In established models this generally leads to unrealistic unbounded development of opinions. Our remedy, also the main idea of this article, is the consideration of relative opinions. As a consequence of considering only relative opinions, the model with $N$ agents effectively evolves in a subspace of dimension $N-1$.

The interaction on which opinion formation is based in the relative opinion model is assumed to be constant over time. This is a major simplification of reality. In order to assess time-dependent interactions, it is expected that one should rely on simulation techniques. Still, a constant influence structure over time may be an approximation of short-term or very long-term opinion formation processes. One could imagine that the influence structure changes incrementally in the short term, meaning that a constant influence structure is still a good proxy for the opinion dynamics. In the long term, it may be reasoned that on average agents stick to a repetitive contact structure with established friends and family. As a remedy for the intermediate term one may consider to approximate the time frame by a collection of short-term intervals that can be approximated by a constant influence pattern.

The idea of modeling relative opinions instead of opinions measured according a predefined scale can be applied to a broad set of established models. In the case of agent based models, the fact that the relative opinion model has an elementary linear update rule may lead to relatively low simulation effort. In the case of models that allow for analytical solutions the idea of relative opinions may extend the options for analytical investigation.

We anticipate that the material presented in this paper is only the tip of the iceberg of the analytical insights that the relative opinion model can provide. A vast body of matrix theory, e.g. theory involving symmetric matrices, positive matrices and block matrices, may be applied to the relative opinion model, possibly providing new insights. As a research direction one may consider the relative opinion model from a stochastic perspective. Another direction is the behavior of the relative opinion model when opinions are treated as an object that is higher than 1-dimensional as considered by many models in the literature \citep{Baldassarri2007,Flache2011b,schweighofer2020}, or when a different equivalence relation is chosen than the one considered in this paper.

From an empirical point of view it is notable that data based on a predefined opinion scale, e.g. a scale from 1 to 10 (the {\it Likert scale}), can be directly inserted in the relative opinion model. One could also aim to validate, in situations where the underlying assumptions are expected to be approximately met, our results for two groups, so as to predict the asymptotic opinion behavior of the population from the intragroup and intergroup dynamics. 

We feel that the idea of relative opinions is a useful additional instrument in the toolkit of theoretical models of opinion formation. Most of all, when this idea is applied to a DeGroot-like or discrete-time Altafini-like model, we obtain a model that unites, in a single simple but general setting, the three major empirical phenomena of polarization, consensus formation, and periodicity.

\appendix

\section{Probabilistic arguments}
The complete Matlab code and datasets used in generating the figures in this article are publicly available through [files are included in the submission and will be made publicly available when the submission is accepted]

\subsection{Notation \& definitions} 
\subsubsection{Notation}
In the sequel we use the following notation:
\begin{itemize}
    \item[$\circ$] Each vector $\boldsymbol{v}$ is a column vector.
    \item[$\circ$] $\lvert M \rvert$ is the determinant of matrix $M$.
    \item[$\circ$] $\Tr{(M)}$ is the trace of matrix $M$.
    \item[$\circ$] With $z=a+b\rm{i}$ denoting a complex number, we denote its absolute value by $\lvert z \rvert$, defined as $\lvert z \rvert = \sqrt{a^2 + b^2}$.
    \item[$\circ$] $I$ is the square identity matrix.
    \item[$\circ$] $\boldsymbol{1}_n$ is the $n$-dimensional all-ones vector.
\end{itemize}

\subsubsection{Definitions}\label{subsubsect_def}
As mentioned in Footnote \ref{foot_singular}, agents with opinion $0$ are excluded from the definition of relative opinions (until this opinion becomes non-zero due to influences by other agents). It is worth noting that $0$-entries of opinion vectors $\boldsymbol{y}^{(t)}$ (notation as in Equation (\ref{eq_ROM_description})) only appear for very specific choices of $A$ and $\boldsymbol{y}^{(0)}$; the settings in which relative opinions are not defined can be seen as  ``pathological'' (i.e., occurring only for very specific instances, where it is in addition noted that the issue is resolved by a small perturbation of the problematic instance).

Another setting where $0$-entries play a role is when relative opinions between two agents \textit{approach} $0$ over time. For example, when agent $1$'s opinion (denoted by $y_1^{(t)}$) is constant and agent $2$'s opinion (denoted by $y_2^{(t)}$) is strictly increasing then -- from the perspective of agent $2$ -- the opinion of agent $1$ (i.e., $y_1^{(t)} / y_2^{(t)}$) approaches $0$ as $t \to \infty$. However -- from the perspective of agent $1$ -- the opinion of agent~$2$ (i.e., $y_2^{(t)} / y_1^{(t)}$) takes arbitrarily large values as $t \to \infty$. This example shows that, in the context of relative opinions ``approaching 0'' is equivalent to ``taking arbitrarily large values''. The concepts of convergence and asymptotic behavior have a meaningful interpretation in situations for which the corresponding values cannot become arbitrarily large. In the definition of convergence below we take a similar approach; we restrict the notion of convergence to vectors with non-zero entries only. Similarly, in the definition of asymptotic behavior we exclude opinion vectors of which its entries take the $0$-value arbitrarily often over time.

\vspace{0.3cm}

\noindent
\textbf{Relative opinions.}\:\:
Consider the $N$-dimensional real-valued vector $\boldsymbol{y}$ with non-zero entries. The relative opinions of $\boldsymbol{y}$ are defined through
\[ y_{ij} = \frac{y_i}{y_j},~\:\:i,j=1, \ldots, N.\]

\vspace{0.3cm}

\noindent
\textbf{Convergence under the equivalence relation $\equiv$.}\:\:
Consider the $N$-dimensional real-valued vectors $\boldsymbol{a}(t)$, $t=0, 1, 2, \ldots$, and $\boldsymbol{b}$. Assume that $\boldsymbol{b}$ has non-zero entries only. Let $a_{ij}(t)$ and $b_{ij}$, $i,j = 1, \ldots, N$ denote the relative opinions corresponding to $\boldsymbol{a}(t)$ and $\boldsymbol{b}$ respectively. If $a_{ij}(t)$ converges to $b_{ij}$ for $i,j = 1, \ldots, N$ then we say that $\boldsymbol{a}(t)$ converges under $\equiv$ to $\boldsymbol{b}$. We write $\lim_{t \to \infty} \boldsymbol{a}(t) \equiv \boldsymbol{b}$.

\vspace{0.3cm}

\noindent
\textbf{Asymptotic behavior.}\:\:
Consider the $N$-dimensional real-valued vectors $\boldsymbol{a}(t)$ and $\boldsymbol{b}(t)$, $t=0, 1, 2, \ldots$. Assume the existence of a $T \in \mathbb{N}$ such that $\boldsymbol{b}(t)$ only has non-zero entries for $t > T$. Let $a_{ij}(t)$ and $b_{ij}(t)$, $i,j = 1, \ldots, N$ denote the relative opinions corresponding to $\boldsymbol{a}(t)$ and $\boldsymbol{b}(t)$. If 
\[\lim_{t\to\infty}|a_{ij}(t) - b_{ij}(t)| \rightarrow0\] for all $i,j = 1, \ldots, N$, then we say that $\boldsymbol{a}(t)$ behaves asymptotically as $\boldsymbol{b}(t)$.

\subsection{General derivations} \label{subsec_app_gen_der}

By $t$ times iterating Equation (\ref{eq_ROM_description}),  we obtain ${\boldsymbol y}^{(t)} = A^t {\boldsymbol y}^{(0)}$. When $A$ is diagonalizable, the eigenvectors of $A$, denoted by $\boldsymbol{v}_1, \boldsymbol{v}_2, \ldots, \boldsymbol{v}_N$ corresponding to eigenvalues $\lambda_1, \lambda_2, \ldots, \lambda_N$, are linearly independent. As a consequence, $V = (\boldsymbol{v}_1 \,|\, \boldsymbol{v}_2 \,| \ldots |\, \boldsymbol{v}_N)$ is invertible. Define $\Lambda$ as the diagonal matrix with $\lambda_1, \lambda_2, \ldots, \lambda_N$ on its diagonal. Then by definition $AV = V \Lambda$, hence $A=V\Lambda V^{-1}$.

We write $\boldsymbol{y}^{(0)}$ in the basis of the eigenvectors: $\boldsymbol{y}^{(0)} = b_1 \boldsymbol{v}_1 + \ldots + b_N \boldsymbol{v}_N = V\boldsymbol{b}$ with $\boldsymbol{b}=(b_1, \ldots, b_N)^{\top} \in \mathbb{C}^N$. In this paper we assume for convenience that all entries of $\boldsymbol{b}$ corresponding to $\boldsymbol{y}^{(0)}$ are non-zero. For the general case, all mathematical reasonings can be replicated with slight adjustments. Without loss of generality, let all eigenvalues of $A$ be indexed in a descending order (i.e., $\lvert \lambda_1 \rvert \leq \lvert \lambda_2 \rvert \leq \ldots \leq \lvert \lambda_N \rvert$). Now:
\begin{equation} \label{standard_form}
    \begin{split}
        \boldsymbol{y}^{(t)} = A^t \boldsymbol{y}^{(0)} = V \Lambda^t V^{-1} V \boldsymbol{b} & = \lambda_1^t b_1 \boldsymbol{v}_1 + \ldots + \lambda^t_N b_N \boldsymbol{v}_N \\
        &= \lvert \lambda_1 \rvert ^t \left(\Big(\frac{\lambda_1}{\lvert \lambda_1 \rvert}\Big)^t b_1 \boldsymbol{v}_1 + \Big(\frac{\lambda_2}{\lvert \lambda_1 \rvert}\Big)^t b_2 \boldsymbol{v}_2 + \ldots + \Big(\frac{\lambda_N}{\lvert \lambda_1 \rvert}\Big)^t b_N \boldsymbol{v}_N\right) \\
        & \equiv \Big(\frac{\lambda_1}{\lvert \lambda_1 \rvert}\Big)^t b_1 \boldsymbol{v}_1 + \Big(\frac{\lambda_2}{\lvert \lambda_1 \rvert}\Big)^t b_2 \boldsymbol{v}_2 + \ldots + \Big(\frac{\lambda_N}{\lvert \lambda_1 \rvert}\Big)^t b_N \boldsymbol{v}_N.
    \end{split}
\end{equation}
When $A$ has $n\leq N$  largest eigenvalues (in absolute terms, that is), $\boldsymbol{y}^{(t)}$ under the equivalence relation behaves asymptotically as \[\Big(\frac{\lambda_1}{\lvert \lambda_1 \rvert}\Big)^t b_1 \boldsymbol{v}_1 + \Big(\frac{\lambda_2}{\lvert \lambda_1 \rvert}\Big)^t b_2 \boldsymbol{v}_2 + \ldots + \Big(\frac{\lambda_n}{\lvert \lambda_1 \rvert}\Big)^t b_n \boldsymbol{v}_n.\] If $n=1$ and $\lambda_1\in{\mathbb R}_+$, then $\boldsymbol{y}^{(t)}$ converges to ${\boldsymbol v}_1$. Likewise, if $n=1$ and $\lambda_1\in{\mathbb R}_-$, but in this case the opinion vector changes sign each time period. If $n=1$ and $\lambda_1\in{\mathbb C}\setminus {\mathbb R}$, then the opinion dynamics are asymptotically periodic according to Theorem \ref{th_comp_conj}. For $n=2,3,\ldots$, the asymptotic behavior of $\boldsymbol{y}^{(t)}$ is a weighted mix of the aforementioned dynamics. When $A$ is a stochastic matrix -- as in the DeGroot model -- one of the largest eigenvalues is $\lambda_1 = 1$ with corresponding eigenvector $\boldsymbol{1}_N$. In \citet{DeGroot} and \citet{ANTHONISSE1977360} conditions are discussed that ascertain $n=1$ and thus consensus formation in the long run, relying on concepts such as irreducibility and aperiodicity.

If $A$ is not diagonalizable, then Equation (\ref{standard_form}) is invalid because $V$ is not invertible. However, a comparable equation can be obtained by extending the reasoning with notions such as Jordan normal form and generalized eigenvectors. The rate of convergence is in this case typically a mixture of exponential and polynomial.

\subsection{Derivations for $N=2$}
The following theorem reveals the asymptotic behavior of the relative opinions as a function of the trace of the matrix $A$ in a two agent population with positive equal attentiveness.

\begin{theorem}
\label{th_k_cap}
Consider relative opinions as desribed by $\boldsymbol{y}^{(t)}$ in equation (\ref{eq_ROM_description}), implying that the equivalence relation holds; $\boldsymbol{y}^{(t)}$ denotes any vector $\gamma \, \boldsymbol{y}^{(t)}$ with $\gamma \in \mathbb{R}_+$. Let $A=\{a_{ij}\}$ be a $2 \times 2$ real-valued $k$-equal attentiveness matrix, with $k \in \mathbb{R}_+$. Let $\boldsymbol{v}_1, \boldsymbol{v}_2$ be the eigenvectors corresponding to the eigenvalues $\lambda_1$, $\lambda_2$. Denote $\boldsymbol{y}^{(\infty)}$ by the limit ($t \to \infty$) of $\boldsymbol{y}^{(t)}$, meaning the limit of the fractions as described by Equation (\ref{eq_relative}), if it exists. An eigenvalue of $A$ is $\lambda_1 = k$ with eigenvector $\boldsymbol{v}_1 = \boldsymbol{1}_2$. The following holds under the equivalence relation:

\begin{enumerate}
    \item[(i)] If $\Tr{(A)} < 0$ and $\boldsymbol{y}^{(0)} \not\equiv \boldsymbol{1}_2$ then $\boldsymbol{y}^{(t)}$ behaves asymptotically with exponential rate as $\boldsymbol{y}^{(t)} \equiv \pm (-1)^t \boldsymbol{v}_2$, with the sign determined by the choice of $\boldsymbol{y}^{(0)}$.
    \item[(ii)] If $0 < \Tr{(A)} < 2k$ and $\boldsymbol{y}^{(0)} \not\equiv \boldsymbol{v}_2$ then $\boldsymbol{y}^{(t)}$ converges exponentially to $\boldsymbol{y}^{(\infty)} \equiv \pm \boldsymbol{1}_2$, with the sign determined by the choice of $\boldsymbol{y}^{(0)}$.
    \item[(iii)] If $\Tr{(A)} > 2k$ and $\boldsymbol{y}^{(0)} \not\equiv \boldsymbol{1}_2$ then $\boldsymbol{y}^{(t)}$ converges exponentially to $\boldsymbol{y}^{(\infty)} \equiv \pm \boldsymbol{v}_2$, with the sign determined by the choice of $\boldsymbol{y}^{(0)}$. If (i) $a_{11} > k$ and $a_{22} > k$ then $\boldsymbol{v}_2$ has entries of opposite signs, if (ii) $a_{11} = k$ or $a_{22} = k$ then $\boldsymbol{v}_2$ has an entry equal to 0, and if (iii) otherwise then $\boldsymbol{v}_2$ has entries of equal signs.
    \item[(iv)] If $\Tr{(A)} = 0$ and $\boldsymbol{y}^{(0)} \not\equiv \boldsymbol{1}_2$ and $\boldsymbol{y}^{(0)} \not\equiv \boldsymbol{v}_2$ then\footnote{The case of $A$ being periodic is included here: $a_{12}=a_{21}=k$.} $\boldsymbol{y}^{(t)}$ behaves as $\boldsymbol{y}^{(t)} \equiv (b_1 \boldsymbol{v}_1 + (-1)^t b_2 \boldsymbol{v}_2)$ for some real constants $b_1, b_2$ that is determined by  the choice of $\boldsymbol{y}^{(0)}$.
    \item[(v)] If $\Tr{(A)} = 2k$, $\boldsymbol{y}^{(0)} \not\equiv \boldsymbol{v}_2$ and not $a_{12}=a_{21}=0$ then\footnote{The case of $A$ with $a_{12}=a_{21}=0$ represents a disjoint population.} $\boldsymbol{y}^{(t)}$ converges with rate $O(t^{-1})$ to $\boldsymbol{y}^{(\infty)} \equiv \pm \boldsymbol{1}_2$, with the sign determined by the choice of $\boldsymbol{y}^{(0)}$.
\end{enumerate}
\end{theorem}

\begin{proof}[Proof of Theorem \ref{th_k_cap}]
Since $A$ is a $k$-equal attentiveness matrix, an eigenvalue of $A$ must be $\lambda_1 = k$ with corresponding eigenvector $\boldsymbol{1}_N$.
Hence, 
\begin{equation} \label{eq_eigenvalue2x2}
    \lambda_{1} = k\: \text{ and }\: \lambda_2 = \Tr{(A)} - k,
\end{equation}
since the sum of the eigenvalues of $A$ equals its trace.
Note that the eigenvalues are real-valued and distinct unless $\Tr{(A)} = 2k$. Thus $A$ is diagonalizable for $\Tr{(A)} \neq 2k$. By the same reasoning as in (\ref{standard_form}):
\begin{equation}\label{eq_2by2}
    \begin{split}
        \boldsymbol{y}^{(t)} &= \lambda_1^t b_1 \boldsymbol{v}_1 + \lambda_2^t b_2 \boldsymbol{v}_2 \\
        & \equiv \Big(\frac{\lambda_1}{\lvert \lambda_1 \rvert}\Big)^t b_1 \boldsymbol{v}_1 + \Big(\frac{\lambda_2}{\lvert \lambda_1 \rvert}\Big)^t b_2 \boldsymbol{v}_2 
         \equiv \Big(\frac{\lambda_1}{\lvert \lambda_2 \rvert}\Big)^t b_1 \boldsymbol{v}_1 + \Big(\frac{\lambda_2}{\lvert \lambda_2 \rvert}\Big)^t b_2 \boldsymbol{v}_2,
    \end{split}
    \end{equation}
with $b_1, b_2$ the scalars of $\boldsymbol{y}^{(0)}$ when expressed in the basis of the eigenvectors of $A$. If $\boldsymbol{y}^{(0)} \equiv \boldsymbol{v}_1 = \boldsymbol{1}_2$, then $b_2 = 0$; if $\boldsymbol{y}^{(0)} \equiv \boldsymbol{v}_2$, then $b_1 = 0$. 
\begin{enumerate}
\item[(i)] If $\Tr{(A)} < 0$ then $\lvert \lambda_2 \rvert > \lvert \lambda_1 \rvert$ with $\lambda_2 < 0$. Equation (\ref{eq_2by2}) then equals \[\lvert \lambda_2\rvert^t \left(\Big(\frac{\lambda_1}{\lvert\lambda_2\rvert}\Big)^t b_1 \boldsymbol{v}_1 + (-1)^t b_2 \boldsymbol{v}_2\right).\] Hence, $\boldsymbol{y}^{(t)}$ behaves asymptotically as $\pm (-1)^t \boldsymbol{v}_2$ if $b_2$ is non-zero. 

\item[(ii)] If $0 < \Tr{(A)} < 2k$ then $\lvert \lambda_1 \rvert > \lvert \lambda_2 \rvert$, clearly $({\lambda_2}/{\lvert \lambda_1 \rvert})^t$ converges exponentially to $0$, hence $\boldsymbol{y}^{(\infty)} \equiv \pm \boldsymbol{v}_1$ if $b_1$ is non-zero or, equivalently, $\boldsymbol{y}^{(0)} \not\equiv \boldsymbol{v}_2$. 

\item[(iii)] If $\Tr{(A)} > 2k$ then $\lambda_2 > \lambda_1 > 0$, thus $({\lambda_1}/{\lambda_2})^t$ converges exponentially to $0$, therefore $\boldsymbol{y}^{(\infty)} \equiv \pm \boldsymbol{v}_2$ when $b_2$ is non-zero. 
Combining Equation (\ref{eq_eigenvalue2x2}) and the definition of eigenvectors yields
\begin{equation}
v_2[2] (k - a_{11}) = -v_2[1] (k - a_{22})
\end{equation}
where $v_2[1], v_2[2]$ are the first and second entries of eigenvector $\boldsymbol{v}_2$. Clearly, $v_2[1]$ and $v_2[2]$ are of opposite signs when $a_{11} > k$ and $a_{22} > k$, and of equal sign otherwise.

\item[(iv)] If $\Tr{(A)} = 0$ then $\lambda_2 = -\lambda_1$, hence $\boldsymbol{y}^{(t)}= \lvert \lambda_1\rvert^t (b_1 \boldsymbol{v}_1 + (-1)^t b_2 \boldsymbol{v}_2)$, and $\boldsymbol{y}^{(t)}$ behaves as $b_1 \boldsymbol{v}_1 + (-1)^tb_2 \boldsymbol{v}_2$.

\item[(v)] If $\Tr{(A)}  = 2k$ then $\lambda_1 = \lambda_2 = k$, $A$ is therefore not necessarily diagonalizable as it does not have distinct eigenvalues. First we prove that the geometric multiplicity of the eigenvalue $k$ is $1$ unless $a_{12}=a_{21}=0$. By definition, the geometric multiplicity of an eigenvalue $\lambda$ is the dimension of the nullspace of $(A-\lambda I)$. Clearly, the $(A-\lambda I)$ matrix is the $0$-matrix if and only if $a_{12}=a_{21}=0$. Hence the dimension of the nullspace of $(A-\lambda I)$ -- or the geometric multiplicity -- is $2$ if and only if $a_{12}=a_{21}=0$. In this situation $A$ represents a disjoint population which is a trivial case. Now, since the geometric multiplicity of $k$ must be larger than $0$ it must be $1$ when $a_{12}=a_{21}=0$ does not hold. For convenience we write $\lambda = \lambda_1 = \lambda_2$. Now, for not $a_{12}=a_{21}=0$ we have
\begin{equation}\label{eq_jordan}
    \begin{split}
        \boldsymbol{y}^{(t)} &= \Big(\lambda^t b_1 + \lambda^{t-1} \binom{t}{1}b_2\Big) \boldsymbol{v}_1 + \lambda^t b_2 \boldsymbol{v}_2 \\
                &= \lambda^t \Big(b_1 \boldsymbol{v}_1 + b_2 \boldsymbol{v}_2\Big) + \lambda^{t-1} t b_2 \boldsymbol{v}_1 = \lambda^t \Big(b_1 {\boldsymbol v}_1 + b_2 \boldsymbol{v}_2 + \frac{t}{\lambda} b_2 \boldsymbol{v}_1\Big) \\
                & \equiv t\Big(\frac{b_1 \boldsymbol{v}_1 +b_2 \boldsymbol{v}_2}{t} + \frac{b_2 \boldsymbol{v}_1}{\lambda}\Big)  \equiv \frac{b_1 \boldsymbol{v}_1 + b_2 \boldsymbol{v}_2}{t} + \frac{b_2 \boldsymbol{v}_1}{\lambda},
    \end{split}
\end{equation}
with $\boldsymbol{v}_2$ a generalized eigenvector.
The first equality in (\ref{eq_jordan}) follows from a known result that involves the Jordan normal form, which is applicable because $\lambda$ is of geometric multiplicity 1. Clearly, $({b_1 \boldsymbol{v}_1 + b_2 \boldsymbol{v}_2})/{t}\to 0$ as $t\to\infty$, hence $\boldsymbol{y}^{(\infty)} \equiv \pm \boldsymbol{v}_1$ if $b_2$ is non-zero, or equivalently $\boldsymbol{y}^{(0)} \not\equiv \boldsymbol{v}_2$.
\end{enumerate}
This completes the proof. \end{proof}

In Theorem \ref{th_comp_conj} we apply a  well-known lemma from linear algebra that describes the basis of real-valued vectors. For the sake of completeness we provide the lemma here.

\begin{lemma}
\label{lm_comp_conj}
Let $\{\boldsymbol{v}_1, \boldsymbol{v}_2, \ldots, \boldsymbol{v}_l, \boldsymbol{w}_1, \bar{\boldsymbol{w}}_{1}, \ldots, \boldsymbol{w}_m, \bar{\boldsymbol{w}}_{m} \}$ be $N = l + 2m$ linearly independent vectors of dimension $N$ with $\boldsymbol{v}_1, \ldots, \boldsymbol{v}_l$ and $\boldsymbol{w}_1, \ldots, \boldsymbol{w}_m$ real-valued and non-real complex-valued vectors respectively. If $\boldsymbol{x}$ is a real-valued vector of dimension $N$ then $\boldsymbol{x}$ can be written in the form:
\begin{equation}
    \boldsymbol{x} = p_1 \boldsymbol{v}_1 + \ldots p_l \boldsymbol{v}_l + q_1 \boldsymbol{w}_1 + \bar{q}_1 \bar{\boldsymbol{w}}_1 + \ldots + q_m \boldsymbol{w}_m + \bar{q}_m \bar{\boldsymbol{w}}_m,
\end{equation}
where $p_1, \ldots, p_l$ are real and $q_1, \ldots, q_m$ are complex numbers.
\end{lemma}

\begin{proof}[Proof of Lemma \ref{lm_comp_conj}]
Since $\boldsymbol{v}_1, \boldsymbol{v}_2, \ldots, \boldsymbol{v}_l$, $\boldsymbol{w}_1, \bar{\boldsymbol{w}}_{1}, \ldots, \boldsymbol{w}_m, \bar{\boldsymbol{w}}_{m}$ are $N$ linearly independent $N$-dimensional  vectors they form a basis of $\mathbb{C}^N$. Hence, unique complex numbers $p_1, \ldots, p_l$, $q_1, \ldots, q_m$, $r_1, \ldots, r_m$ exist such that
\begin{equation}
    \boldsymbol{x} = p_1 \boldsymbol{v}_1 + \ldots p_l \boldsymbol{v}_l + q_1 \boldsymbol{w}_1 + r_1 \bar{\boldsymbol{w}}_1 + \ldots + q_m \boldsymbol{w}_m + r_m \bar{\boldsymbol{w}}_m .
\end{equation}
Since $\boldsymbol{x}$ and $\boldsymbol{v}_1, \ldots, \boldsymbol{v}_l$ are real-valued, the complex conjugate of the previous equation is
\begin{equation}
    \boldsymbol{x} = \bar{p}_1 \boldsymbol{v}_1 + \ldots \bar{p}_l \boldsymbol{v}_l + \bar{q}_1 \bar{\boldsymbol{w}}_1 + \bar{r}_1 \boldsymbol{w}_1 + \ldots + \bar{q}_m \bar{\boldsymbol{w}}_m + \bar{r}_m \boldsymbol{w}_m.
\end{equation}
Since this decomposition is unique, the factors corresponding to the vectors must be identical. Hence, $p_{i}$, $i = 1, 2, \ldots, l$ are real numbers and $r_j = \bar{q}_j$, $j = 1, \ldots, m$.
\end{proof}

\begin{theorem}
\label{th_comp_conj}
Consider $\boldsymbol{y}^{(t)}$ in Equation (\ref{eq_ROM_description}), let $A$ be a real-valued diagonalizable matrix. Assume $\boldsymbol{y}^{(0)}$ has only non-zero complex-valued coordinates $b_1, b_2, \ldots, b_N$ in the basis of eigenvectors, $\boldsymbol{v}_1, \ldots, \boldsymbol{v}_N$ of $A$. Let non-real and complex-valued $\lambda_1$ and its conjugate be the absolute largest eigenvalues. Denote by $\theta$ the argument of $\lambda_1$. If the entries of $\operatorname{Re}\big( e^{i \theta t}b_1 \boldsymbol{v}_1\big)$ only take non-zero values for $t=0,1,2, \ldots$ then $\boldsymbol{y}^{(t)}$ behaves asymptotically as
\begin{equation}
    \boldsymbol{y}^{(t)} \equiv \operatorname{Re}\big( e^{i \theta t}b_1 \boldsymbol{v}_1\big).
\end{equation}
\end{theorem}

\begin{proof}[Proof of Theorem \ref{th_comp_conj}]
Since $A$ is diagonalizable, Equation (\ref{standard_form}) holds. As $A$ is real-valued, the complex conjugate of the eigenvalue equation $A\boldsymbol{v}_1 = \lambda_1 \boldsymbol{v}_1$ is $A\bar{\boldsymbol{v}}_1 = \bar{\lambda}_1 \bar{\boldsymbol{v}}_1$, thus $\bar{\lambda}_1$ is also an eigenvalue of $A$ with eigenvector $\bar{\boldsymbol{v}}_1$ with $\lvert \lambda_1 \rvert = \lvert \bar{\lambda}_1 \rvert$. Let $\lambda_2$ in Equation (\ref{standard_form}) be the complex conjugate of $\lambda_1$, then
\[
    \begin{split}
        \boldsymbol{y}^{(t)} & = \lambda_1^t b_1 \boldsymbol{v}_1 + \bar{\lambda}_1^t \bar{b}_1 \bar{\boldsymbol{v}}_1 + \ldots + \lambda^t_N b_N \boldsymbol{v}_N \\
        &= \lvert \lambda_1 \rvert^t \big( \cos(\theta t) + i\sin(\theta t) \big) b_1 \boldsymbol{v}_1 + \lvert \lambda_1 \rvert^t \big( \cos(\theta t) - i\sin(\theta t) \big) \bar{b}_1 \bar{\boldsymbol{v}}_1 + \ldots + \lambda^t_N b_N \boldsymbol{v}_N \\
        &= \lvert \lambda_1 \rvert^t \Big[2 \operatorname{Re} \Big( (\cos(\theta t) + i\sin(\theta t))b_1 \boldsymbol{v}_1 \Big) + \Big(\frac{\lambda_3}{\lvert \lambda_1 \rvert}\Big)^t b_3 \boldsymbol{v}_3 + \ldots + \Big(\frac{\lambda_N}{\lvert\lambda_1\rvert}\Big)^t b_N \boldsymbol{v}_N \Big] \\
        & \equiv 2 \operatorname{Re}\Big(\big(\cos(\theta t) + i\sin(\theta t)\big)b_1 \boldsymbol{v}_1\Big) + \Big(\frac{\lambda_3}{\lvert \lambda_1 \rvert}\Big)^t b_3 \boldsymbol{v}_3 + \ldots + \Big(\frac{\lambda_N}{\lvert\lambda_1\rvert}\Big)^t b_N \boldsymbol{v}_N, \\
    \end{split}
\]
where the $\bar{b}_1$ in the first equality follows from Lemma $\ref{lm_comp_conj}$. The second and third equalities are obtained by expressing $\lambda_1$ in polar coordinates with $0 < \theta < 2\pi$ and using that the addition of a complex vector with its conjugate is two times the real part of the complex vector, respectively. The result follows from Euler's formula and the assumption that entries of $\operatorname{Re}\big( e^{i \theta t}b_1 \boldsymbol{v}_1\big)$ do not take non-zero values for $t=0,1,2,\ldots$ (where we remark that the assumption is discussed in detail in the two paragraphs immediately after this proof). 
\end{proof}

We like to note that $\boldsymbol{y}^{(t)}$ as in Theorem \ref{th_comp_conj} may behave differently from $\operatorname{Re}( e^{i \theta t}b_1 \boldsymbol{v}_1)$ when $\operatorname{Re}( e^{i \theta t}b_1 \boldsymbol{v}_1)$ \textit{can} consistently attain $0$-entries for $t=0, 1, 2, \ldots$. It is stressed, though, that such scenarios can be seen as ``pathological'', in that very specific parameter values have to be chosen, and that in addition the issue is resolved when slightly perturbing these parameters.

As an illustration, consider the situation where $A$ and $\boldsymbol{y}^{(0)}$ are chosen such that two entries of $\boldsymbol{y}^{(t)}$ (notation as in the proof of Theorem \ref{th_comp_conj}) are equivalent to functions of the form $\cos(\theta t + \phi_1) + e^{-t}$ and $\cos(\theta t + \phi_2) + e^{-2t}$ ($\theta$ is the argument of $\lambda_1$, $\phi_1$ and $\phi_2$ are arguments of different entries of $\boldsymbol{v}_1$ and $e^{-t}$, $e^{-2t}$ follow from specific values of the expressions $(\lambda_j / |\lambda_1|)^t$, $j = 3, \ldots, N$). Further, when parameters are chosen such that $\phi_1 = \phi_2 = \pi / 2$, then the relative opinions equal \[\Bigg[\frac{\cos(\theta t + \pi / 2) + e^{-2t}}{\cos(\theta t + \pi / 2) + e^{-t}},\: \frac{\cos(\theta t + \pi / 2) + e^{-t}}{\cos(\theta t + \pi / 2) + e^{-2t}}\Bigg]. \]

Now, in the case that $\theta = \pi$ the relative opinions equal $[e^{-t},\:e^{t}]$ since $\cos(\theta t + \pi /2)=0$. However, when $\theta$ is a value such that $\theta t \neq \pi k$ for all $k = 1, 2, \ldots$ and $t=0,1,2, \ldots$, then evidently $\cos(\theta t + \pi / 2) \neq 0$ for all $t=0,1,2, \ldots$, so  that the relative opinions  converge to $[1,1]$ as $t \to \infty$. Thus, for very specific parameters  the relative opinions behave as $[e^{-t},\:e^{t}]$, whereas  in ``normal'' situations the relative opinions converge under the equivalence relation to $[1,1]$. In Theorem \ref{th_comp_conj} it is therefore assumed that $\operatorname{Re}( e^{i \theta t}b_1 \boldsymbol{v}_1)$ attains only non-zero entries for $t=0, 1, 2, \ldots$, being a sufficient condition to exclude the ``pathological'' settings as described above.

\begin{proposition}
\label{prop_comp_22}
Let $A=\{a_{ij}\}$ be a $2 \times 2$ real-valued matrix. Then $A$ has non-real and complex eigenvalues if and only if 
\begin{equation}
    (a_{11} - a_{22})^2 < -4a_{12}a_{21}.
\end{equation}
\end{proposition}

\begin{proof}[Proof of Theorem \ref{prop_comp_22}]
The determinant $D$ of the quadratic characteristic polynomial is 
\begin{equation}
    \begin{split} 
        D & = (a_{11} + a_{22})^2 - 4(a_{11}a_{22} - a_{12}a_{21}) \\
        &= a_{11}^2 +a_{22}^2 - 2 a_{11}a_{22} + 4 a_{12}a_{21} \\
        & = (a_{11} - a_{22})^2 + 4 a_{21} a_{12}. \\
    \end{split}
\end{equation}
The eigenvalues of $A$ are non-real and complex-valued if and only if $D < 0$.
\end{proof}

\subsection{Derivations for a two group population}
Before showing the proof of Theorem \ref{th_22ev} which substantiates various statements in Section \ref{sect_groups}, we present for the sake of completeness the well-known Schur complement and a useful lemma on the determinant of matrices.

\begin{lemma}[\citet{schur1917}\label{schur_det}]
Let the square matrix $A$ be partitioned in the block matrices $E$, $F$, $G$, $H$ with square matrices on the diagonal. Let the matrix $E$ be invertible, then
\begin{equation}\label{schur_det1}
\left| A \right| = \left|
\begin{array}{cc}
E&F\\
G&H\\
\end{array}
\right|=
\left|E\right| \left| H  - G E^{-1} F \right|
\end{equation}
\end{lemma}
\begin{proof}[Proof of Lemma \ref{schur_det}]
We may write 
\begin{equation}
A = \left(
\begin{array}{cc}
E&F\\
G&H\\
\end{array}
\right) = \left(
\begin{array}{cc}
E&0\\
G&I\\
\end{array}
\right) 
\left(
\begin{array}{cc}
I&E^{-1}F\\
0& H-G E^{-1} F\\
\end{array}
\right);
\end{equation}
taking determinants we obtain (\ref{schur_det1}).
\end{proof}

\begin{lemma}[\citet{DING20071223}\label{mat_det_lem}]
Consider an invertible matrix $M$ and vectors $\boldsymbol{u}$ and $\boldsymbol{v}$. Then
\begin{equation}
    |M + \boldsymbol{u} \boldsymbol{v}^{\top}| = (1 + \boldsymbol{v}^{\top} M^{-1}\boldsymbol{u})|M|.
\end{equation}
\end{lemma}
\begin{proof}[Proof of Lemma \ref{mat_det_lem}]
For the special case $M = I$ we have
\begin{equation}
    \left(
        \begin{array}{cc}
        I&0\\
        \boldsymbol{v}^{\top}&1\\
        \end{array}
    \right) 
    \left(
        \begin{array}{cc}
        I + \boldsymbol{u} \boldsymbol{v}^{\top} & \boldsymbol{u}\\
        0 & 1\\
        \end{array}
    \right) 
    \left(
        \begin{array}{cc}
        I & 0\\
        -\boldsymbol{v}^{\top} & 1\\
        \end{array}
    \right) = \left(
        \begin{array}{cc}
        I & \boldsymbol{u}\\
        0 & \boldsymbol{v}^{\top} \boldsymbol{u} + 1\\
        \end{array}
    \right); 
\end{equation}
taking determinants we obtain $|I + \boldsymbol{u}\boldsymbol{v}^{\top}| = (1 + \boldsymbol{v}^{\top} \boldsymbol{u})$. For the general case we have
\begin{equation}
    |M + \boldsymbol{u} \boldsymbol{v}^{\top}| = |M|  \cdot |I + M^{-1}\boldsymbol{u} \boldsymbol{v}^{\top}| = |M| (1 + \boldsymbol{v}^{\top} M^{-1} \boldsymbol{u}).
\end{equation} 
This proves the claim.
\end{proof}

The following theorem substantiates the claims on the population's asymptotic dynamics as a result of its groups' asymptotic dynamics.

\begin{theorem}
\label{th_22ev}
Let $C = (c_{ij})$ be a $2 \times 2$, $P_{11}$ a $m_1 \times m_1$ and $P_{22}$ a $m_2 \times m_2$ complex matrix. Assume $P_{12} = \boldsymbol{1}_{m_1} \boldsymbol{v}^{\top}$, $\boldsymbol{v} \in \mathbb{C}^{m_2}$, a $m_1 \times m_2$ complex matrix with all columns linear dependent, and $P_{21}$ a $m_2 \times m_1$ complex matrix. Let $P_{ij}, 1 \leq i,j \leq 2$ be 1-equal attentiveness matrices and denote the eigenvalues of $c_{11}P_{11}$ and $c_{22}P_{22}$ by $\alpha_1 = c_{11}, \alpha_2, \ldots, \alpha_{m_1}$ and $\beta_1 = c_{22}, \beta_2, \ldots, \beta_{m_2}$ respectively. Then the set of eigenvalues of the matrix\footnote{Matrix $A$ may be referred as the so-called Khatri-Rao product of $C$ and $P$, where $P$ is the block matrix consisting of blocks $P_{ij}, 1 \leq i,j \leq 2$.}
\begin{equation}
A = 
\left(
\begin{array}{cc}
c_{11}P_{11}&c_{12} \boldsymbol{1}_{m_1} \boldsymbol{v}^{\top}\\
c_{21}P_{21}&c_{22}P_{22}\\
\end{array}
\right)
\end{equation}
equals the eigenvalues of $C$ and $\alpha_2, \ldots, \alpha_{m_1}$, $\beta_2, \ldots, \beta_{m_2}$.
\end{theorem}

\begin{proof}[Proof of theorem \ref{th_22ev}]
The ideas underlying this proof are similar to the ones as described in \citet{OUELLETTE1981187}. The characteristic polynomial $p(\lambda)$ is
\begin{equation}
\begin{split}
    p(\lambda) & = 
        \left| 
            \begin{array}{cc}
                c_{11}P_{11} - \lambda I &c_{12} \boldsymbol{1}_{m_1} \boldsymbol{v}^{\top}\\
                c_{21}P_{21}&c_{22}P_{22} - \lambda I\\
            \end{array}
        \right|\\
    & = |c_{11}P_{11} - \lambda I| \cdot |c_{22}P_{22} - \lambda I - c_{21} P_{21} (c_{11}P_{11} - \lambda I)^{-1} c_{12} \boldsymbol{1}_{m_1} \boldsymbol{v}^{\top}|,
\end{split}
\end{equation}
using Lemma (\ref{schur_det}) for $\lambda \in \mathbb{C}$ not eigenvalue of $c_{11} P_{11}$. 

We have $(c_{11}P_{11} - \lambda I) \boldsymbol{1}_{m_1} = (c_{11} - \lambda) \boldsymbol{1}_{m_1}$, hence $(c_{11}P_{11} - \lambda I)^{-1} \boldsymbol{1}_{m_1} = (c_{11} - \lambda)^{-1} \boldsymbol{1}_{m_1}$. So,
\begin{equation}
\begin{split}
    p(\lambda) & = |c_{11}P_{11} - \lambda I| \cdot \left|c_{22}P_{22} -  \lambda I - \frac{c_{21} c_{12} \boldsymbol{1}_{m_2} \boldsymbol{v}^{\top}} {c_{11} - \lambda}\right| \\
    & = |c_{11}P_{11} - \lambda I| \cdot |c_{22}P_{22} - \lambda I| \cdot \left| I 
    - \frac{c_{21} c_{12} \boldsymbol{1}_{m_2} \boldsymbol{v}^{\top}} {(c_{11} - \lambda)(c_{22} - \lambda)} \right|
\end{split}
\end{equation}
for $\lambda$ not eigenvalue of $c_{22}P_{22}$, using $P_{21} \boldsymbol{1}_{m_1} = \boldsymbol{1}_{m_2}$. From Lemma \ref{mat_det_lem} and $\boldsymbol{v}^{\top} \boldsymbol{1}_{m_2} = 1$ we obtain
\begin{align}
    p(\lambda) & = |c_{11}P_{11} - \lambda I| \cdot |c_{22}P_{22} - \lambda I| \cdot \left( 1 
    - \frac{c_{21} c_{12}} {(c_{11} - \lambda)(c_{22} - \lambda)} \right)\\
    & = \left(\prod_{i=2}^{m_{1}} (\alpha_i - \lambda)\right) \left( \prod_{j=2}^{m_2} (\beta_{j} - \lambda) \right) [(c_{11} - \lambda) (c_{22} - \lambda) - c_{12}c_{21}].\label{th_22ev1}
\end{align}
We have shown Equation (\ref{th_22ev1}) for $\lambda$ not eigenvalue of $c_{11}P_{11}$ and $c_{22}P_{22}$, but since each complex polynomial is continuous in $\mathbb{C}$, Equation (\ref{th_22ev1}) must hold for all $\lambda \in \mathbb{C}.$ 
\end{proof}

Theorem \ref{th_22ev} suggests that the asymptotic development of relative opinions in a group-structured population as defined in Subsection \ref{subsect_group_transfer} -- under non-trivial conditions -- can be fully determined by the eigenvalues of the intragroup influences and the influences between groups as determined by the matrix $C$. When the (assumed) unique largest eigenvalue of $A$ is an eigenvalue of $C$, it is easy to verify that the corresponding eigenvector of $A$ is a ``higher dimension'' version of  the corresponding eigenvector of $C$. Hence, in a way, asymptotic dynamics of $C$ are transferred to the entire group-structured population. In fact, when the asymptotic dynamics of the stand-alone population with driving matrix $C$ would lead to polarization, for example under one of the conditions as stated in Theorem \ref{th_k_cap}, then the entire population would show consensus formation within the groups and simultaneously polarization between the groups in the long run. The situation in which the dynamics of the entire population can be decomposed occurs under a non-trivial condition. Namely, for one of the groups, its members must be influenced identically by the members of the other group ($P_{12} = \boldsymbol{y}_{m_1} \boldsymbol{v}^{\top}$), while the other group can be influenced in any way ($P_{21}$ can be any matrix). 
\subsection{Methodology behind the generation of Figure \ref{fig_Change}}\label{subsect_Fig1Generation}
In order to produce a figure that depicts two subpopulations with a desired asymptotic behavior, matrix $A$ is considered under the conditions of Theorem \ref{th_22ev}. If the (in absolute terms, that is) largest eigenvalue arises from the matrix $C$, then it is ascertained that the asymptotic behavior of $C$ is transferred to the entire population. Matrix $C$ is chosen to be an $1.2$-equal attentiveness matrix; the choice of $1.2$, a number larger than $1$, is to ensure a positive drift in the opinion development of the population. $C$ is chosen such that it represents a mini-population with a stubborn and an open-minded agent when viewed as a stand-alone system. The parameters of $C$ are such that asymptotic polarization at one side of the sign follows from Theorem~\ref{th_k_cap}. The entries of the $1$-equal attentiveness matrices $P_{11}$, $P_{12}$, $P_{21}$, and $P_{22}$ are uniformly sampled from the interval $[-0.2,0.8]$, and each row is corrected by a multiplication factor such that each row sums up to $1$. Intuitively, the positive drift when choosing the entries  makes sure that $P_{11}$ and $P_{22}$ as stand-alone systems move to consensus asymptotically, or, mathematically, that the dominant eigenvalue is $1$.

\subsection{Sensitivity analyses of Theorem \ref{th_22ev}}\label{subsect_SensA}
Theorem \ref{th_22ev} considers the situation of a two-group population wherein agents of one group are identically influenced by agents of the other group. In this situation the qualitative asymptotic opinion development equals the qualitative asymptotic development of the first group, second group, or the interaction of the groups as described by $C$, as stand-alone populations. 

It is expected in real-life that agents of one group are never identically influenced by agents of the other group. Also, agents within one group do not always have identical attentiveness degrees. Sensitivity analyses are performed showing to what degree the outcomes of Theorem \ref{th_22ev} still hold when we slightly deviate from the mentioned conditions.

\begin{figure}[ht]
\begin{subfigure}[t]{0.312\textwidth}
\centering
\includegraphics[width=0.95\linewidth]{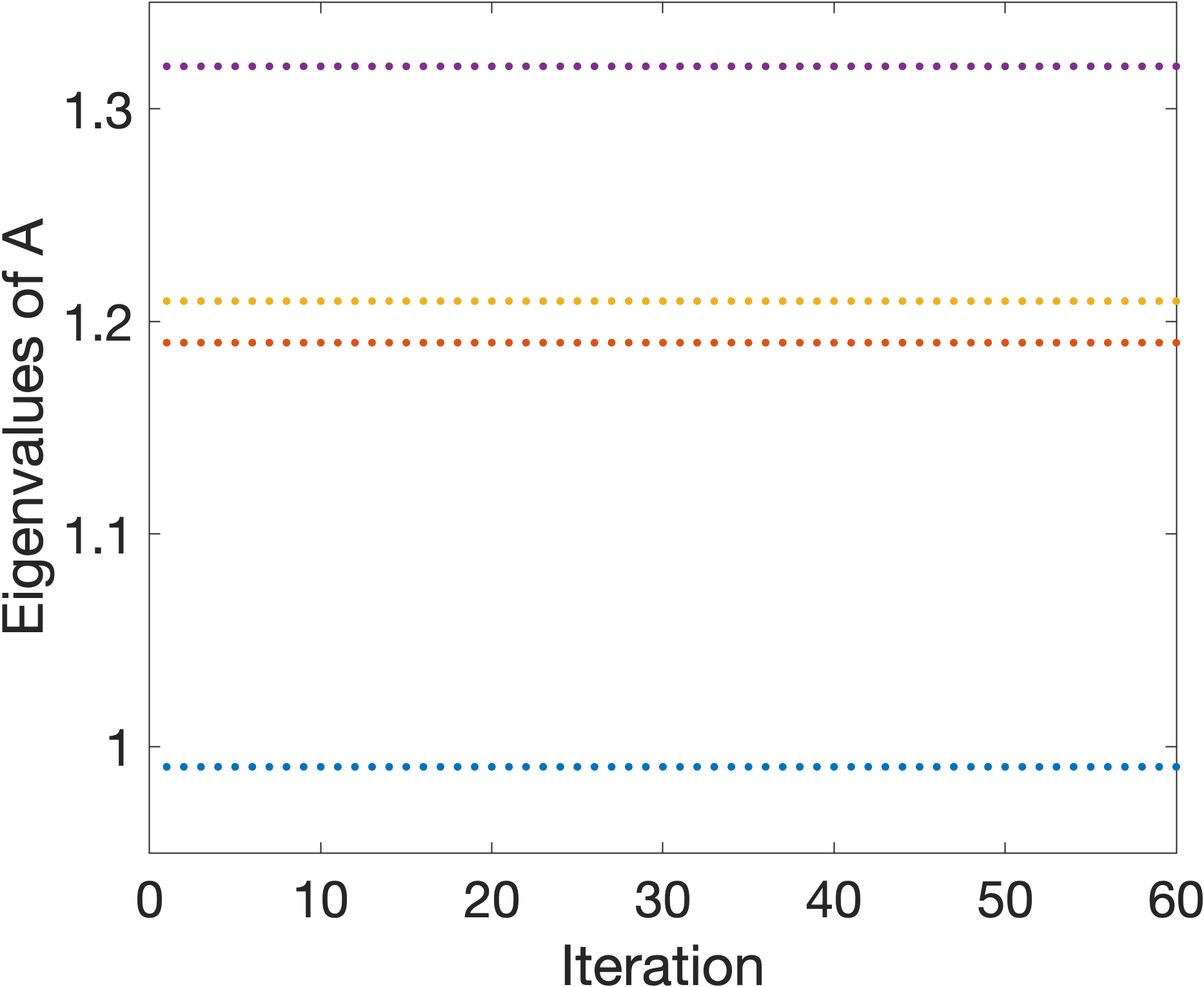} 
\caption{$P_{12}$ satisfies the condition of Theorem \ref{th_22ev}, $P_{12}$ and $P_{21}$ are uniformly sampled around $0$ at each iteration.}
\label{fig_SensANoNoise}
\end{subfigure}\:\:\:
\begin{subfigure}[t]{0.312\textwidth}
\centering
\includegraphics[width=0.95\linewidth]{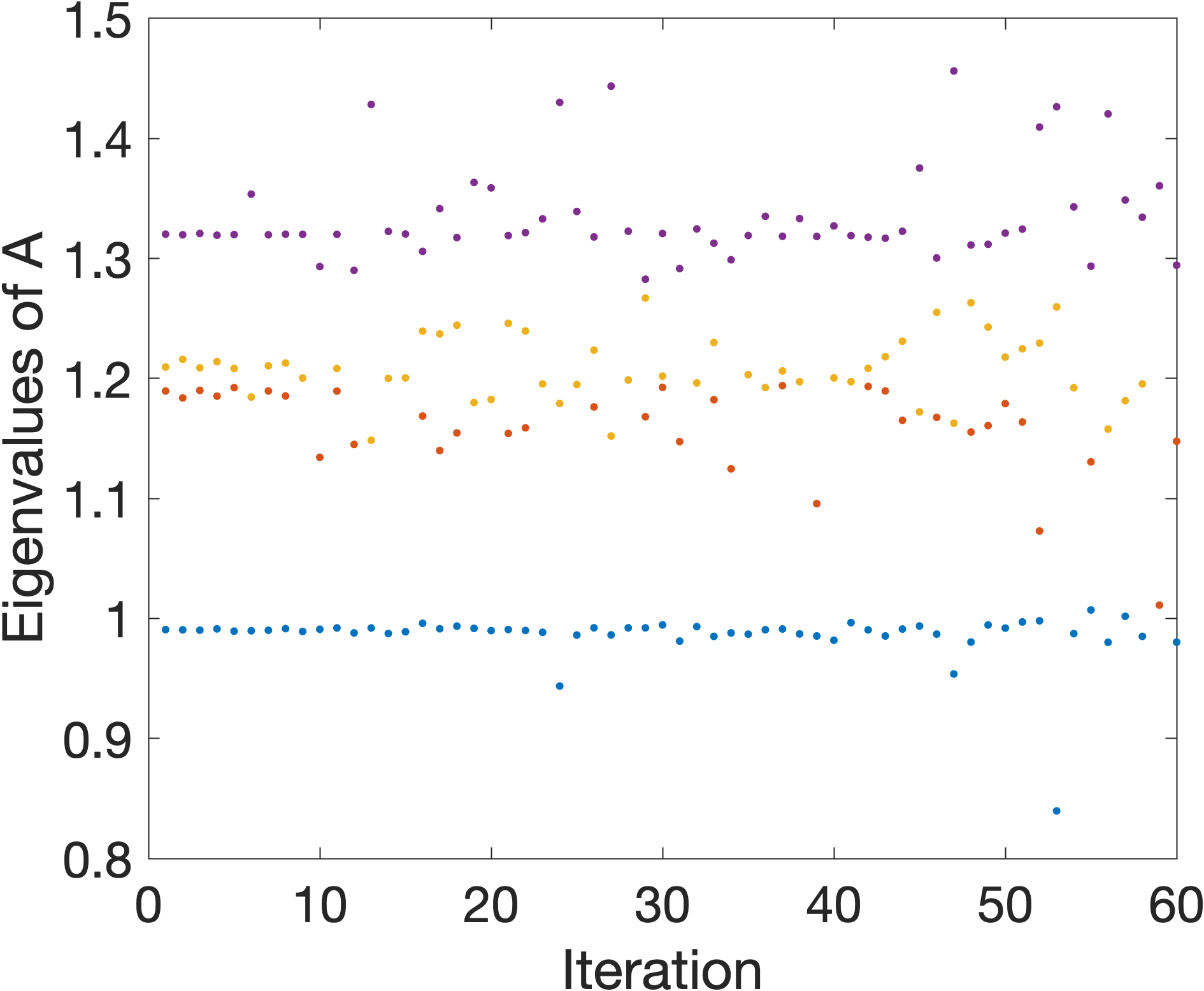}
\caption{Add noise to each entry of $P_{12}$; noise is uniformly sampled from $[-0.02j,0.02j]$ where $j$ represents the $j^{\rm th}$ iteration.}
\label{fig_SensAA12Noise}
\end{subfigure}\:\:\:
\begin{subfigure}[t]{0.312\textwidth}
\centering
\includegraphics[width=0.95\linewidth]{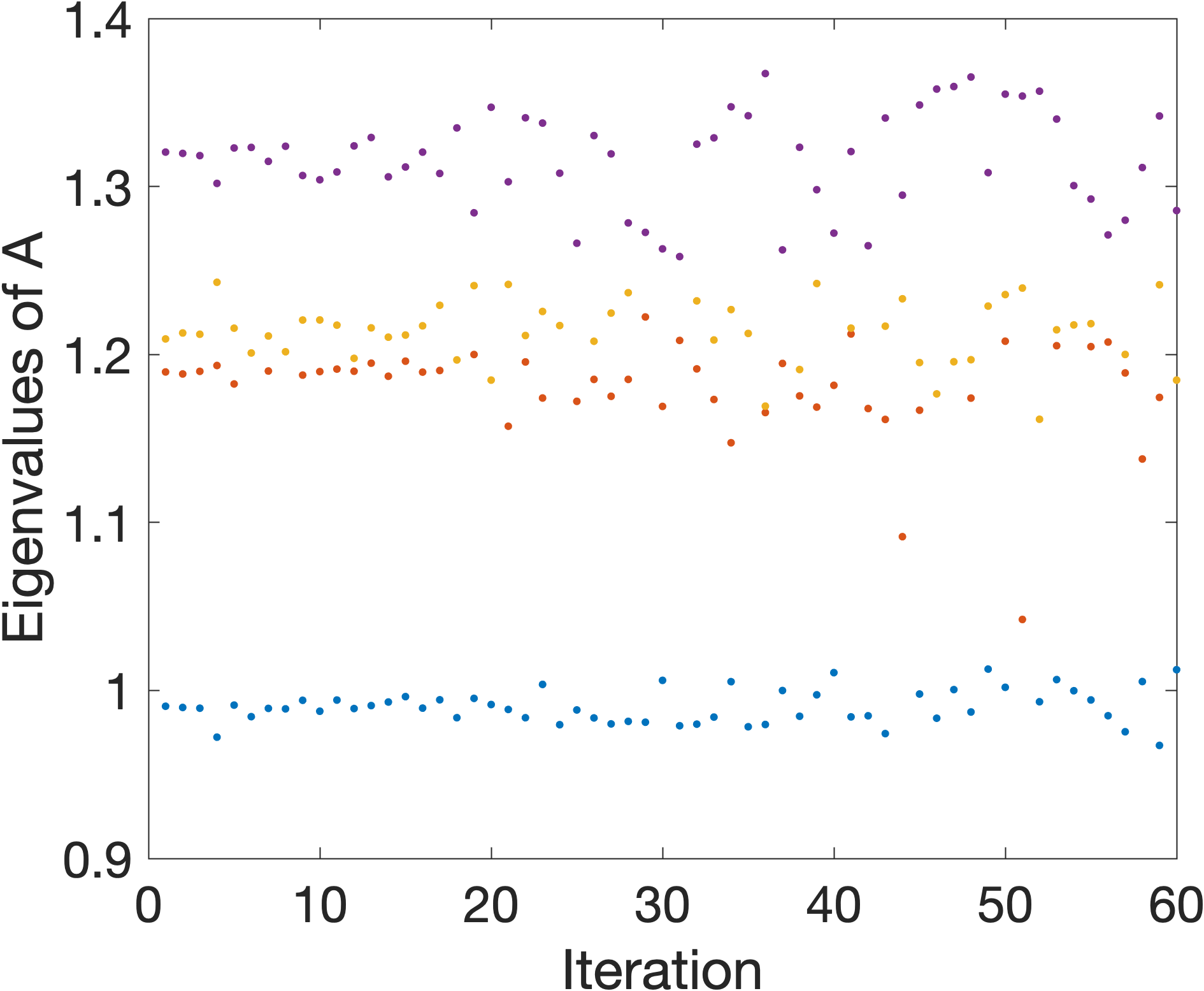}
\caption{Multiplying each row of the inter- and intragroup matrices by a noise factor sampled from $[1-0.0005j,1+0.0005j]$ where $j$ represents the $j^{\rm th}$ iteration.}
\label{fig_SensAAttentNoise}
\end{subfigure}

\caption{Three types of sensitivity analyses of Theorem \ref{th_22ev}}

\label{fig_SensA}
\end{figure}

The results of the sensitivity analyses are depicted in Figure \ref{fig_SensA}. At every iteration step,  $P_{11}$ and $P_{22}$ remain constant, while entries of $P_{12}$ and $P_{21}$ are chosen uniformly between $[-0.5, 0.5]$ before correction by a multiplication factor to enforce equal attentiveness. Before performing the sensitivity analyses, $P_{12}$ satisfies the condition of Theorem \ref{th_22ev}, i.e., it has identical values in each column, but possibly different values between columns. Figure \ref{fig_SensANoNoise} shows how the eigenvalues of $A$ remain constant even while $P_{12}$ and $P_{21}$ are sampled at each iteration step, as predicted by Theorem \ref{th_22ev}. 

Figure \ref{fig_SensAA12Noise} suggests that Theorem \ref{th_22ev} still roughly holds, even when agents of one group are not (exactly) identically influenced by members of the other group. It shows the simulation results in the case that to each entry of $P_{12}$, a noise is added that is uniformly sampled from $[-0.02j,0.02j]$, with $j$ the $j^{\rm th}$ iteration step; this means that by iteration $60$, noise parameters are added that are uniformly sampled from $[-1.2,1.2]$. Note that after the addition of noise, $P_{12}$ is neither a $1$-equal attentiveness matrix nor a matrix with identical values in each column anymore.

Figure \ref{fig_SensAAttentNoise} suggests that essentially the qualitative outcomes of Theorem \ref{th_22ev} still hold, even when agents belonging to the same group do not have identical degrees of attentiveness towards their own and the other group, but the degrees of attentiveness match only approximately. Each row of each inter- and intragroup matrix is multiplied by a uniformly sampled factor from $[1-0.0005j,1+0.0005j]$, where $j$ is the $j^{\rm th}$ iteration step. Thus by iteration step $60$, attentiveness degrees varies between $0.97$ and $1.03$. Note that after the multiplication by a noise factor, the inter- and intragroup matrices typically are not equal attentiveness matrices anymore, and in addition $P_{12}$ is not a matrix with identical values in each column.

\end{document}